\documentclass[10pt,letterpaper]{article}

\usepackage[aboveskip=1pt,labelfont=bf,labelsep=period,justification=raggedright,singlelinecheck=off]{caption}
\usepackage[top=0.85in,left=2.75in,footskip=0.75in]{geometry}
\usepackage[usenames,dvipsnames]{color}
\usepackage{lastpage,fancyhdr,graphicx}
\usepackage{algpseudocode,algorithm}
\usepackage{textcomp,marvosym}
\usepackage{amsmath,amssymb}
\usepackage{nameref,hyperref}
\usepackage[utf8x]{inputenc}
\usepackage[table]{xcolor}
\usepackage{changepage}
\usepackage{microtype}
\usepackage{epstopdf}
\usepackage{xspace}
\usepackage{lipsum}
\usepackage{cases}
\usepackage{array}
\usepackage{soul}
\usepackage{cite}

\raggedright
\DisableLigatures[f]{encoding = *, family = * }

\newcolumntype{+}{!{\vrule width 2pt}}

\newlength\savedwidth

\newcommand\thickhline{\noalign{\global\savedwidth\arrayrulewidth\global\arrayrulewidth 2pt}%
\hline
\noalign{\global\arrayrulewidth\savedwidth}}

\makeatletter
\renewcommand{\@biblabel}[1]{\quad#1.}
\makeatother

\fancyheadoffset[L]{2.25in}
\fancyfootoffset[L]{2.25in}
\DeclareGraphicsExtensions{.pdf,.eps,.jpg,.png}

\definecolor{black}{RGB}{0,0,0}
\definecolor{gray}{RGB}{160,160,160}
\definecolor{blue}{RGB}{102,184,228}
\definecolor{orange}{RGB}{235,138,68}
\definecolor{green}{RGB}{142,186,67}

\algrenewcommand{\algorithmiccomment}[1]{\hfill\{#1\}}

\newcommand{\mean}[1]{\langle#1\rangle}

\newcommand{\figref}[1]{Fig~\ref{fig:#1}\xspace}
\newcommand{\tblref}[1]{Table~\ref{tbl:#1}\xspace}
\renewcommand{\eqref}[1]{Eq~(\ref{eq:#1})\xspace}
\renewcommand{\algref}[1]{Algorithm~\ref{alg:#1}\xspace}

\newcommand{\metref}[0]{Materials and methods\xspace}
\newcommand{\resref}[0]{Results and discussion\xspace}

\DeclareRobustCommand{\hlc}[1]{{\sethlcolor{white}\hl{#1}}}
\DeclareRobustCommand{\hlg}[1]{{\sethlcolor{white}\hl{#1}}}

\begin{document}\vspace*{0.2in}

\begin{flushleft}
{\Large\textbf\newline{War pact model of shrinking networks}}\newline
\\
Luka Naglić\textsuperscript{1}, Lovro Šubelj\textsuperscript{2*}
\\\bigskip
\hlc{\textsuperscript{1} University of Zagreb, Faculty of Science, Zagreb, Croatia}\\
\hlc{\textsuperscript{2} University of Ljubljana, Faculty of Computer and Information Science, Ljubljana, Slovenia}
\\\bigskip
\hlc{* Corresponding author}\\\smallskip
\hlc{E-mail: lovro.subelj@fri.uni-lj.si (L\v{S})}
\end{flushleft}

% % % % % % % % % % % % %
%
%			ABSTRACT
%
% % % % % % % % % % % % %

\section*{Abstract}

Many real systems can be described by a set of interacting entities forming a complex network. To some surprise, these have been shown to share a number of structural properties regardless of their type or origin. It is thus of vital importance to design simple and intuitive models that can explain their intrinsic structure and dynamics. These can\hlg{, for instance,} be used to study networks analytically or to construct networks not observed in real life. Most models proposed in the literature are of two types. A model can be either static, where edges are added between a fixed set of nodes according to some predefined rule, or evolving, where the number of nodes or edges \hlg{increases} over time. However, some real networks do not grow \hlg{but rather} shrink, meaning that the number of nodes or edges \hlg{decreases} over time. We here propose a simple model of shrinking networks called the war pact model. We show that networks generated in such \hlg{a way} exhibit common structural properties of real networks. Furthermore, compared to classical models, these resemble international trade, correlates of war, Bitcoin transactions and other networks \hlg{more closely}. Network shrinking may therefore represent a reasonable explanation of the evolution of some networks and greater emphasis should be put on such models in the future.

% \linenumbers

% % % % % % % % % % % % %
%
%			INTRODUCTION
%
% % % % % % % % % % % % %

\section*{Introduction}

The most natural representation of many real complex systems is a network of nodes connected by edges also called a graph in discrete mathematics. Despite being a very simplistic representation, networks have \hlg{given us a better understanding of} complex real-world phenomena such as epidemic spreading of diseases~\cite{PV01a,PV01b}, small-worlds of human society~\cite{Mil67,WS98}, mobility and navigation~\cite{Kle00,SGMB12}, emergence of complex organization~\cite{BA99,SFMV03}, robustness and controllability of manmade technology~\cite{AJB00,LSB11b}, and the structure of science~\cite{FBBEHMPRSUVWWB18}, to name just a few examples. Indeed, the networks have proven to be an invaluable tool \hlg{for} data analysis in the last two decades~\cite{Bar12}.

One of the key reasons for the successes mentioned above is the realization that real networks share a number of structural properties regardless of their type or origin. For instance, most real networks exhibit \hlg{a scale-free} structure like power-law node degree distribution~\cite{Pri76,BA99}, short distances between the nodes called the small-world structure~\cite{Mil67,WS98}, resilience or robustness to targeted attacks~\cite{AJB00}, pronounced mixing between the nodes~\cite{New02,New03b}, \hlg{a distinctive} mesoscopic network structure~\cite{BE00,NG04}, characteristic node connection patterns~\cite{MSIKCA02,PCJ04} and \hlg{a key} position or centrality of a small number of nodes~\cite{Fre79,BP98}. It is therefore a common belief that real networks form according to some shared rules or principles giving rise to these complex structures.

The network science literature is abundant with generative models of network formation that try to explain their intrinsic structure and dynamics. Most network models are static\hlg{, meaning} that edges are added between a fixed set of nodes according to some predefined rule. These include the simplest Erd\H{o}s-R\'{e}nyi random graphs~\cite{ER59}, and somewhat more realistic configuration~\cite{NSW01}, hierarchical~\cite{CMN08}, geometric~\cite{PCJ04} and optimization~\cite{DBCBK07} graphs that can already explain some non-trivial properties of real networks. Moreover, stochastic block models~\cite{HLL83} can generate networks with \hlg{an arbitrary} mesoscopic structure. However, greater insights into the structure and dynamics of real networks were actually obtained with evolving network models where the number of nodes or edges \hlg{increases} over time. Most well-known examples of evolving models are undoubtedly \hlg{the Price} cumulative advantage model~\cite{Pri76}, \hlg{the Barab\'{a}si-Albert} scale-free networks~\cite{BA99} and \hlg{the copying} network model~\cite{KKRRT99}.

On the other hand, some real networks do not grow but \hlg{rather} shrink, meaning that the number of nodes or edges \hlg{decreases} over time. \hlc{Apart from a few exceptions,} such as~\cite{KNB08}, shrinking network models have been largely neglected in the literature~\cite{DM02}. To fill this gap, we here propose a simple model of shrinking networks called the war pact model. The model starts with some fixed number of edges and \hlg{the} maximal possible number of nodes, \hlc{hence the initial seed network is a perfect matching.} The nodes are then iteratively merged until the desired number of nodes is obtained. We show that networks generated by the war pact model match \hlg{the} most common properties of real networks. More importantly, the model provides an intuitive explanation of the evolution of diverse real networks. The paper therefore puts forth an intriguing question whether growing or shrinking models explain the evolution of real networks \hlg{better}.

% % % % % % % % % % % % %
%
%			METHODS
%
% % % % % % % % % % % % %

\section*{\metref}

The present section describes networks, models and methods used in the paper. We start with a detailed description of the war pact model and its implementation. Next, we introduce four real networks used for empirical validation of the model and alternative random graph models used for comparison. Finally, we review two information-theoretic measures used for comparing networks or graphs.

\subsection*{War pact model}

\begin{figure}[t]
	\includegraphics[width=\linewidth]{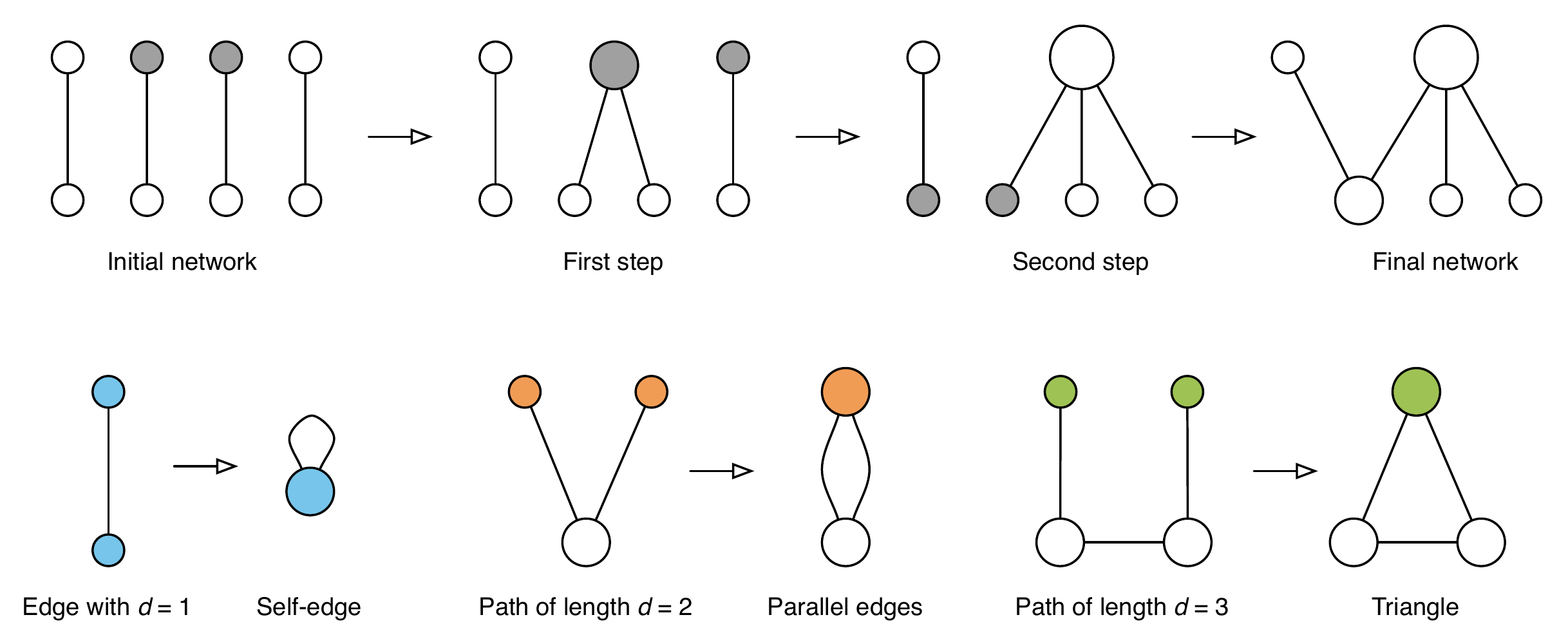}\\%
	\caption{{\bf War pact model.} (\emph{top}) Realization of the war pact model network with $n=5$ nodes and $m=4$ edges. The nodes selected for merging \hlg{in} each step are shown with filled ellipses, while the sizes of the nodes are proportional to their degree $k$. (\emph{bottom}) Examples of the merging procedure for nodes at different \hlg{distances} $d$.}
	\label{fig:warpact}
\end{figure}

The top row in~\figref{warpact} shows a diagram of a particular realization of the war pact model. \hlc{The model starts with an initial seed network which is a perfect matching of nodes with some predefined number of edges.} The model then iteratively merges the nodes until one obtains a network with the desired number of nodes. Note that the number of edges stays fixed during the evolution of the model, while the number of nodes decreases by one \hlg{in} each step. The nodes to be merged \hlg{in} each step can be selected uniformly at random, preferentially according to their degrees or using some other selection rule.

More formally, let $n$ and $m$ be the desired number of nodes and edges, where $2m\geq n$. The model starts with $m$ edges connecting $2m$ nodes as in~\figref{warpact}. \hlg{In} each step, the model merges two nodes $i$ and $j$ into a newly added node $k$ by first replacing nodes $i$ and $j$ with node $k$ and then connecting the neighbors of nodes $i$ and $j$ to node $k$. The model proceeds for $2m-n$ steps when the number of nodes \hlg{equals} $n$.

As shown in the bottom row in~\figref{warpact}, the model can generate \hlg{a rich} local structure depending on the distance $d$ between the nodes being merged. Merging nodes at distance $d=1$ (i.e.\ an edge) creates a self-edge, which is not allowed, merging nodes at distance $d=2$ creates parallel edges and thus a multigraph, while merging nodes at distance $d=3$ creates a triangle resulting in non-trivial network clustering~\cite{WS98}. In general, merging nodes at distance $d$ creates a cycle on $d$ nodes.

The war pact model is free \hlg{from} parameters. Nevertheless, one \hlg{can still freely choose} the strategy \hlg{of selecting} the nodes to be merged \hlg{in} each step and also the initial state of the model. In the \hlg{letter case}, initializing the model with \hlc{a perfect matching} as above is somewhat artificial and not realistic in practice. However, as we show in the~\resref section, the particular choice of the model initialization has no apparent effect on the final structure of the generated networks. For this reason, the model is initialized with \hlc{a perfect matching} unless explicitly stated otherwise.

On the other hand, the particular choice of the node selection rule can have a profound effect on the structure of the generated networks. Therefore, we consider four different node selection rules that proved reasonable in practice. In particular, the two nodes to be merged can be selected uniformly at random among all nodes (denoted RR model) or preferentially according to their degrees (KK model). Hence, a node is selected with the probability proportional to $k$, where $k$ is the current degree of the node. Finally, we also consider two mixed rules where the first node is selected with the probability proportional to its degree $k$, while the second node is selected uniformly at random (KR model) or with the probability proportional to its inverse degree $k^{-1}$ (KI model). Other possible rules either do not generate realistic networks or we could not find an intuitive explanation for such a model.

\begin{figure}[t]
	\includegraphics[width=0.33\linewidth]{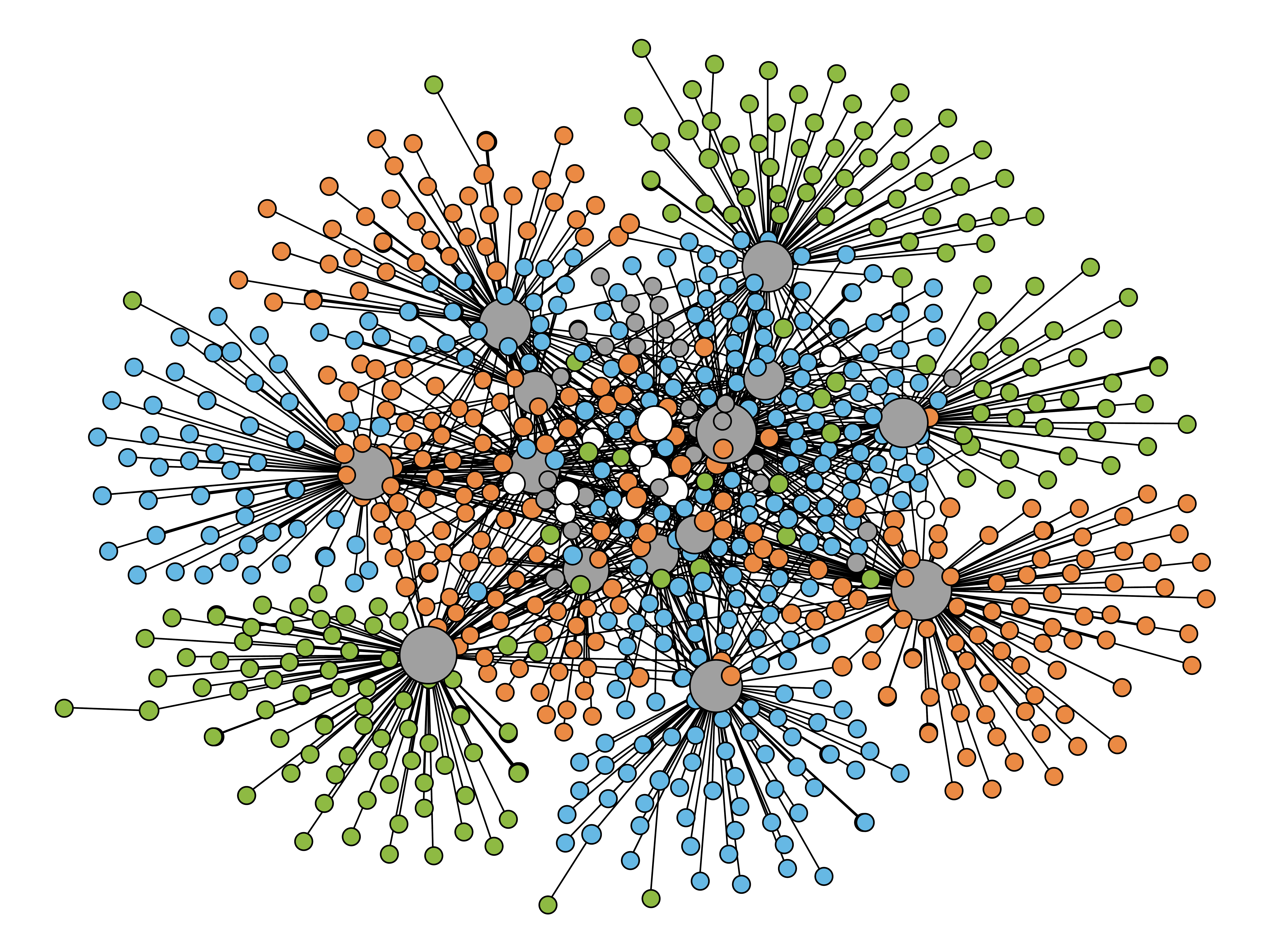}%
	\includegraphics[width=0.33\linewidth]{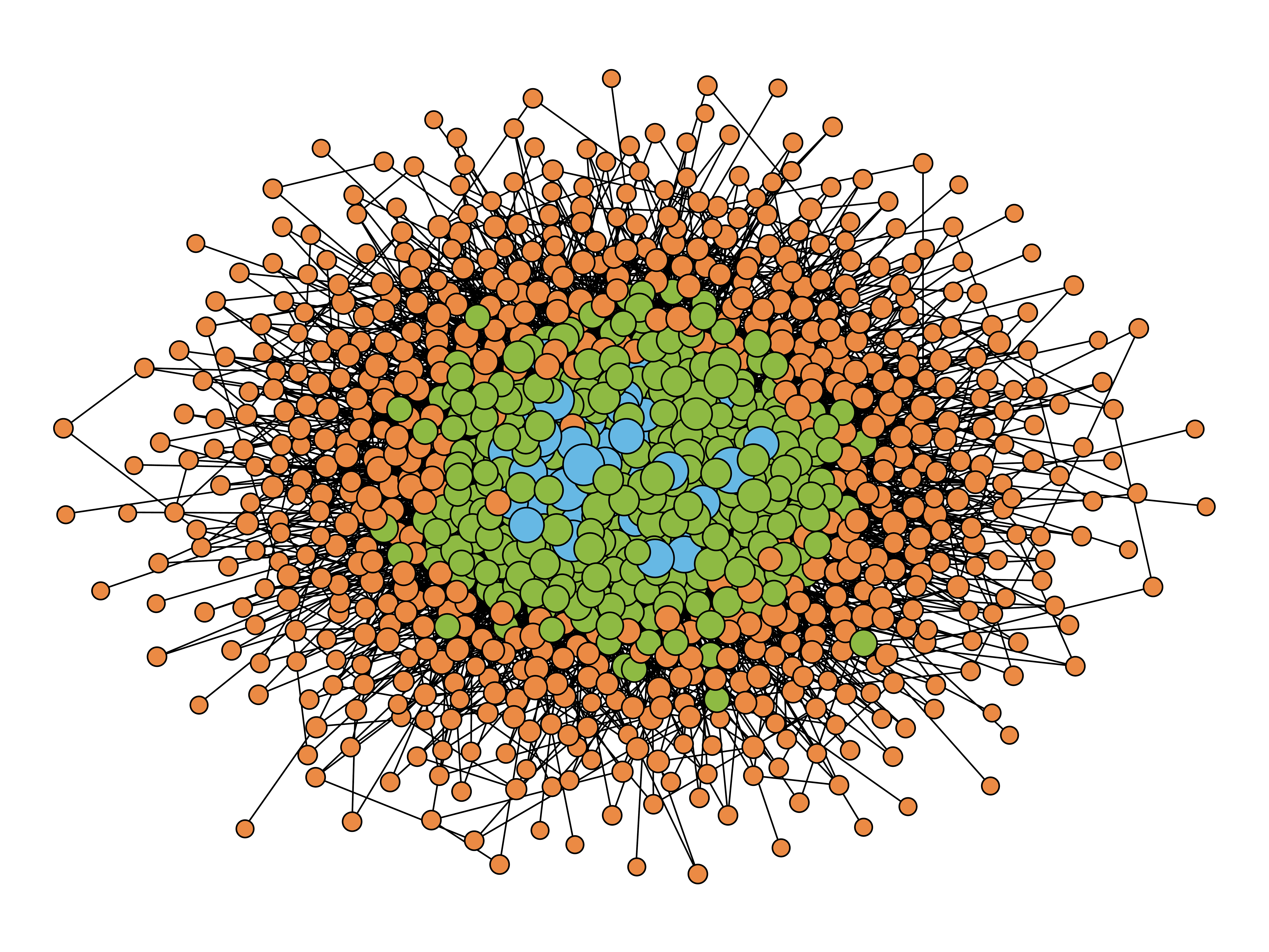}%
	\includegraphics[width=0.33\linewidth]{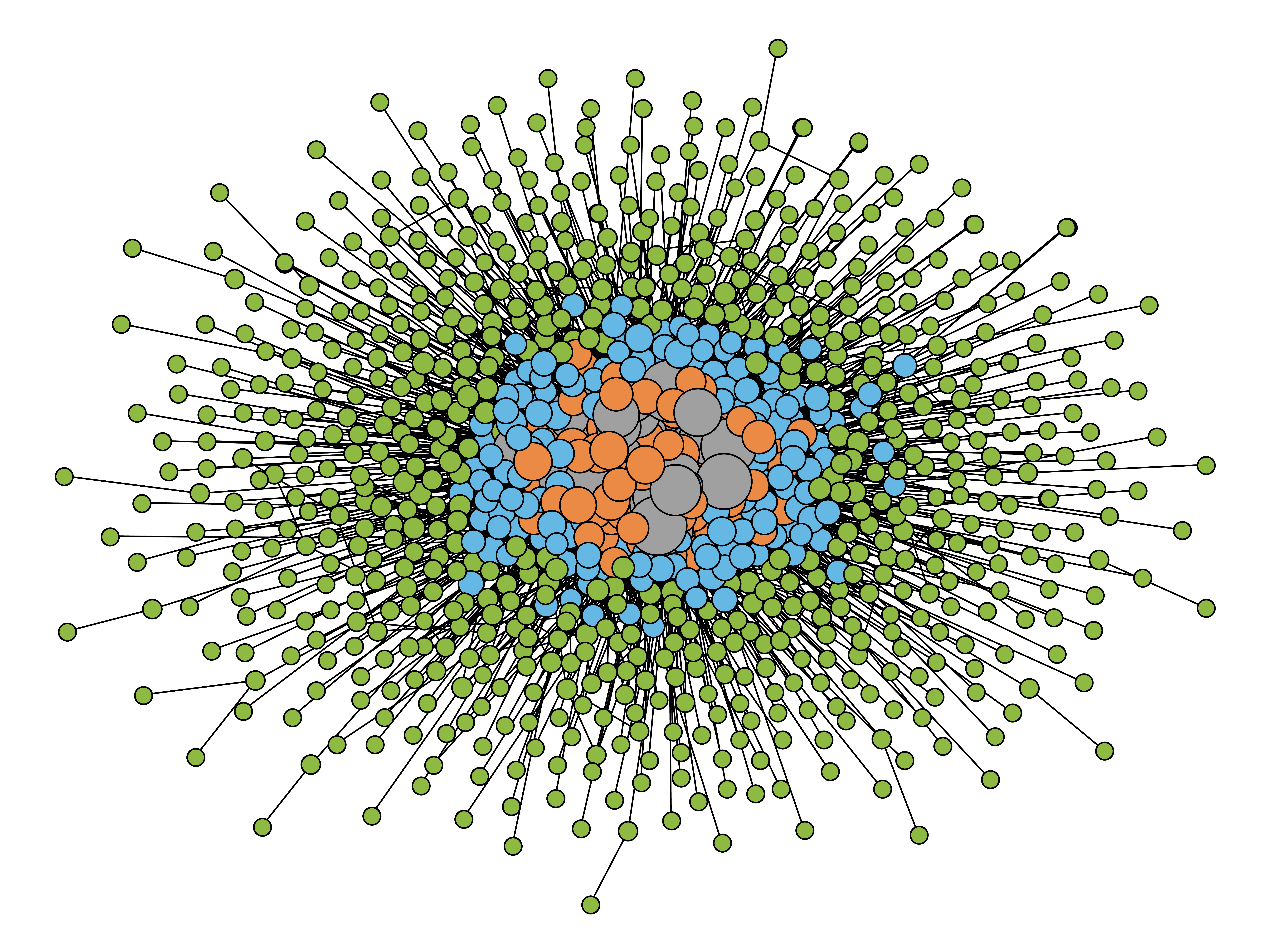}\\%
	\caption{{\bf Layouts of war pact networks.} Wiring diagrams of the largest connected components of the war pact model networks with $n=1\,000$ nodes and the average degree $\mean{k}=10$. The sizes of the nodes are proportional to their degree $k$, while the colors of the nodes show the clusters revealed with stochastic blockmodeling. The layouts were computed with the Large Graph Layout~\cite{ADWM04}.}
	\label{fig:layouts}
\end{figure}

For a visual representation, \figref{layouts} shows layouts of three particular realizations of the war pact model networks. In all three cases, the first node is selected with the probability proportional to its degree $k$, whereas the second node is selected with the probability proportional to its degree $k$, inverse degree $k^{-1}$ or uniformly at random (KK, KI and KR models, respectively). Notice that clusters revealed with Bayesian stochastic blockmodeling~\cite{Pei19} show diverse mesoscopic \hlg{structures} of these networks ranging from hub and spokes arrangements to \hlg{a community} and core-periphery structure.

The implementation of the war pact model is relatively straightforward using \hlg{the} hash map $H$ as shown in~\algref{model}. Each node \hlc{of graph $G$} is represented by its hash value $h\in H$ initialized as $H(i)=i$ for each \hlc{node index} $i=1,\dots,2m$ (lines 4-5). \hlc{Note that each hash value $h\in H$ corresponds to a unique node in graph $G$ and each node has a unique hash value.} Merging two nodes $H(i)$ and $H(j)$ then merely requires unifying their hash values as $H(i)=H(j)$ \hlc{and updating graph $G$ accordingly} (lines 12-13). \hlg{For choosing} a node uniformly at random, one selects a random hash value $h\in H$ (line 9), while \hlg{if choosing} a node with the probability proportional to its degree, one \hlc{selects} the hash value $H(i)$ of a randomly selected \hlc{node index} $i\in[1,2m]$ (line 10). The pseudocode in~\algref{model} assumes that \hlc{graph $G$} is initialized with \hlc{a perfect matching of nodes} (line 6) and ensures that no self-edges are created during the evolution of the model (line 11). Note that, in practice, one should use a disjoint-set data structure instead of a hash map to ensure \hlg{a near-constant} time complexity of all operations.

\begin{algorithm}[t]
\caption{\label{alg:model}War pact model}
\begin{algorithmic}[1]
\Require nodes $n$ and edges $m$
\Ensure graph $G$
\State $H\gets$ empty map \Comment{\hlc{Define empty map representing nodes.}}
\State $G\gets$ empty graph \Comment{\hlc{Define empty war pact model graph.}}
\For{$i\in[1,m]$}
	\State $H(i)\gets i$ and $H(m+i)\gets m+i$ \Comment{\hlc{Map nodes' indices to their hashes.}}
	\State add nodes $H(i)$ and $H(m+i)$ to $G$ \Comment{\hlc{Add nodes (i.e. hashes) to graph.}}
	\State add edge $\{H(i),H(m+i)\}$ to $G$ \Comment{\hlc{Create perfect matching of nodes.}}
\EndFor
\While{$G$ has $>n$ nodes}
	\State $h\gets \Call{Random}{H}$ \Comment{\hlc{Select random hash (i.e. random node).}}
	\State $i\gets \Call{Random}{[1,2m]}$ \Comment{\hlc{Select random index (i.e. node by degree).}}
	\If{$h\neq H(i)$ {\bf and} edge $\{h,H(i)\}\notin G$}
		\State merge nodes $h$ and $H(i)$ in $G$ \Comment{\hlc{Merge selected nodes by rewiring edges.}}
		\State $H(i)\gets h$ \Comment{\hlc{Unify hashes of selected nodes.}}
	\EndIf
\EndWhile
\State \Return $G$
\end{algorithmic}
\end{algorithm}

\subsection*{Networks and models}

For empirical validation of the war pact model, we consider four real networks of different \hlg{types and origins}. The networks represent international trade consisting of the strongest food import and export relations between countries from the Food and Agricultural Organization of the United Nations~\cite{DNAL15}, historical records of international wars, non-military conflicts, border disputes and other disagreements between \hlc{national alliances} during 1996 collected by the Correlates of War project~\cite{DM15b}, % recorded protein-protein interactions of the parasite {\it Plasmodium falciparum} collected from the BioGRID repository in late 2016~\cite{SBRBBT06} 
Bitcoin transactions between the most active users (i.e.\ clusters of coappearing input addresses) between 2012 and 2013 parsed from the public ledger~\cite{KCSPV14}, and the Internet map at the level of autonomous systems on the first day of 1998 reconstructed from the University of Oregon Route Views project~\cite{LKF07}. Networks are represented with undirected graphs with self-edges and isolated nodes removed.

\tblref{statistics} shows the standard statistics of the \hlg{analysed} networks. These are the number of nodes $n$ and edges $m$, the average node degree $\mean{k}=2m/n$, the fraction of nodes in the largest connected component \emph{LCC}, the average node clustering coefficient $\mean{C}=\frac{1}{n}\sum_iC_i$~\cite{WS98} with the clustering coefficient of node $i$ defined as $C_i=\frac{2t_i}{k_i(k_i-1)}$, where $t_i$ is the number of triangles including node $i$ and $k_i>1$ is its degree, the average distance between the nodes $\mean{d}=\frac{2}{n(n-1)}\sum_{i<j}d_{ij}$, where $d_{ij}$ is the number of edges in the shortest paths between nodes $i$ and $j$, the maximal distance or diameter $d_{max}=\max_{i<j}d_{ij}$, the node degree mixing coefficient $r$~\cite{New02} defined as the Pearson's correlation coefficient of the degrees of connected nodes and the modularity of network community structure $Q=\frac{1}{2m}\sum_{ij}(A_{ij}-\frac{k_ik_j}{2m})\,\delta(c_i,c_j)$~\cite{GN02}, where $A$ is the network adjacency matrix, $c_i$ is the community label of node $i$ and $\delta$ is the Kronecker delta. The modularity $Q$ is reported as the average over $100$ runs of the Leiden algorithm~\cite{TWV19}.

\begin{table}[t]
	\begin{adjustwidth}{-2.25in}{0in}
	\centering
	\caption{\hlc{\bf Statistics of real networks.} Standard statistics of real networks analysed in the paper.}\vskip8pt
	\begin{tabular}{|l+r|r|r|r|r|r|c|r|r|} \hline
		{\bf Network} & \multicolumn{1}{c|}{$n$} & \multicolumn{1}{c|}{$m$} & \multicolumn{1}{c|}{\emph{LCC}} & \multicolumn{1}{c|}{$\mean{k}$} & \multicolumn{1}{c|}{$\mean{C}$} & \multicolumn{1}{c|}{$\mean{d}$} & \multicolumn{1}{c|}{$d_{max}$} & \multicolumn{1}{c|}{$r$} & \multicolumn{1}{c|}{$Q$} \\\thickhline
		Correlates of war & \hlc{$41$} & \hlc{$54$} & \hlc{$87.8\%$} & \hlc{$2.63$} & \hlc{$0.28$} & \hlc{$2.58$} & \hlc{$8$} & \hlc{$-0.29$} & \hlc{$0.60$} \\\hline
		International trade & $130$ & $3\,730$ & $100.0\%$ & $57.38$ & $0.50$ & $2.24$ & $5$ & $-0.07$ & $0.21$ \\\hline
		% Protein interactions & $1\,206$ & $2\,545$ & $96.0\%$ & $4.22$ & $0.02$ & $3.90$ & $10$ & $0.09$ & $0.54$ \\\hline
		Bitcoin transactions & $1\,288$ & $6\,236$ & $98.8\%$ & $9.68$ & $0.33$ & $2.83$ & $9$ & $-0.28$ & $0.39$ \\\hline
		Autonomous systems & $3\,213$ & $11\,248$ & $100.0\%$ & $7.00$ & $0.18$ & $3.77$ & $9$ & $-0.22$ & $0.64$ \\\hline
	\end{tabular}
	\label{tbl:statistics}
	\end{adjustwidth}
\end{table}

The war pact model is compared against three classical random graph models. The first is the Erd\H{o}s-R\'{e}nyi random graph model~\cite{ER59}, where an edge is put between each pair of $n$ nodes with \hlg{a probability of} $\mean{k}/(n-1)$. Next is the Barab\'{a}si-Albert scale-free model~\cite{BA99}, where $n$ nodes are added one at a time and each forms $\mean{k}/2$ edges while preferentially linking to high degree nodes. The model generates networks with a scale-free degree distribution $p_k\sim k^{-\gamma}$~\cite{BA99,FFF99}, where $\gamma$ is the power-law exponent. Finally, we consider the Watts-Strogatz small-world model~\cite{WS98}, where a fraction of edges of a regular ring lattice is randomly rewired. The model generates networks with \hlg{a high} clustering coefficient $\mean{C}\gg 0$ and \hlg{a short} average distance between the nodes $\mean{d}\simeq \log_{\mean{k}} n$.

\subsection*{Network comparison}

We adopt two recently proposed measures for comparing networks or graphs. These are the simplified $D$-measure~\cite{SCDPMR17} and the portrait divergence~\cite{BB18,BBSA08}. Both are principled information-theoretic measures that can be used to compare arbitrary graphs and do not require that the two graphs being compared are defined on the same set of nodes. \hlc{Both measures compare graphs by quantifying differences among the distances between the nodes of the graphs as defined below.}

Let $d_{ij}(G)$ denote the distance between nodes $i$ and $j$ in an undirected graph $G$ and $d_{max}(G)$ the maximal distance or diameter, $d_{max}(G)=\max_{i<j}d_{ij}(G)$. Next, \hlc{let $D_{id}(G)$ be the fraction of nodes at distance $d$ from node $i$}, $d=0,\dots,d_{max}(G)$,
\begin{eqnarray}
	D_{id}(G) & = & \frac{1}{n}\sum_{j=1}^n\mathcal{I}({d_{ij}(G)=d)}, \nonumber
\end{eqnarray}
where $\mathcal{I}$ is the indicator function. Finally, \hlc{let $\mathcal{D}(G)$ be the average of vectors $D_i(G)$ over all nodes in $G$, therefore}
\begin{eqnarray}
	\mathcal{D}_d(G) & = & \frac{1}{n}\sum_{i=1}^nD_{id}(G). \nonumber
\end{eqnarray}
The simplified $D$-measure~\cite{SCDPMR17} measuring \hlc{the dissimilarity between graphs $G$ and $G'$} is then defined as 
\begin{eqnarray}
	\label{eq:dmeasure}
	D(G,G') & = & \frac{1}{2}\sqrt{\frac{\mathcal{J}(\mathcal{D}(G),\mathcal{D}(G'))}{\log2}} + \frac{1}{2}\left|\sqrt{\mathcal{N}(G)}-\sqrt{\mathcal{N}(G')}\right|,
\end{eqnarray}
where $\mathcal{N}(G)$ is the so-called node dispersion of graph $G$,
\begin{eqnarray}
	\mathcal{N}(G) & = & \frac{\mathcal{J}(D_1(G),\dots,D_n(G))}{\log(d_{max}(G)+1)}, \nonumber
\end{eqnarray}
and $\mathcal{J}$ is the Jensen-Shannon divergence.

\hlc{The first term of}~\eqref{dmeasure} \hlc{compares graphs through averaged distances between the nodes and thus captures global differences between the graphs. The second term further compares graphs through the heterogeneity of the nodes and how each particular node is connected throughout the graph. It thus captures local differences between the graphs. It was empirically shown that the measure returns non-zero values only for non-isomorphic graphs}~\cite{SCDPMR17}.
 
In the case of the complete $D$-measure, \eqref{dmeasure} also includes \hlc{the third} term measuring the \hlc{dissimilarity} between node centralities in graphs $G$ and $G'$, and their complements. Since the latter are computationally prohibitive for sparse graphs, and only strictly necessary to distinguish highly regular graphs, we here avoid the additional term without significant precision loss~\cite{SCDPMR17}.

Furthermore, \hlc{let $P_{kd}(G)$ be the number of nodes that have $k$ nodes at distance $d$}, $d=0,\dots,d_{max}(G)$,
\begin{eqnarray}
	P_{kd}(G) & = & \sum_{i=1}^n\mathcal{I}(nD_{id}(G)=k), \nonumber
\end{eqnarray}
while other details are the same as before. \hlc{$P(G)$ is called the portrait of graph $G$, which is invariant under graph isomorphism}~\cite{BBSA08}. The portrait divergence~\cite{BB18} measuring the distance between graphs $G$ and $G'$ is then defined as
\begin{eqnarray}
	\label{eq:portrait}
	P(G,G') & = & \mathcal{J}(\mathcal{P}(G),\mathcal{P}(G')),
\end{eqnarray}
where \hlc{$\mathcal{J}$ is the Jensen-Shannon divergence and}
\begin{eqnarray}
	\label{eq:P}
	\mathcal{P}_{kd}(G) & = & \frac{1}{n}P_{kd}(G)\frac{1}{\sum_cn_c^2}\sum_{k'=0}^nk'P_{k'd}(G).
\end{eqnarray}
Here, $n_c$ is the number of nodes in \hlg{the} connected component $c$ and the sum in the denominator goes through all connected components of $G$. \hlc{The right part of}~\eqref{P} \hlc{equals the probability that two randomly chosen nodes are at distance $d$, while the left part further demands that one of these two nodes has exactly $k$ nodes at distance $d$.} The portrait divergence in~\eqref{portrait} has a number of desirable properties for comparing graphs \hlg{thoroughly} described in~\cite{BB18}.

% % % % % % % % % % % % %
%
%			RESULTS
%
% % % % % % % % % % % % %

\section*{\resref}

\hlg{This section presents an empirical validation of the war pact model.} First, we characterize \hlg{the} statistical properties of the networks generated by different variants of the model. Next, we study the model evolution by analyzing networks with growing number of nodes or edges. Finally, we compare the war pact model against classical random graph models and clarify the intuition behind the model for various real networks.

\subsection*{War pact networks}

\begin{figure}[t]
	\includegraphics[width=0.29\linewidth]{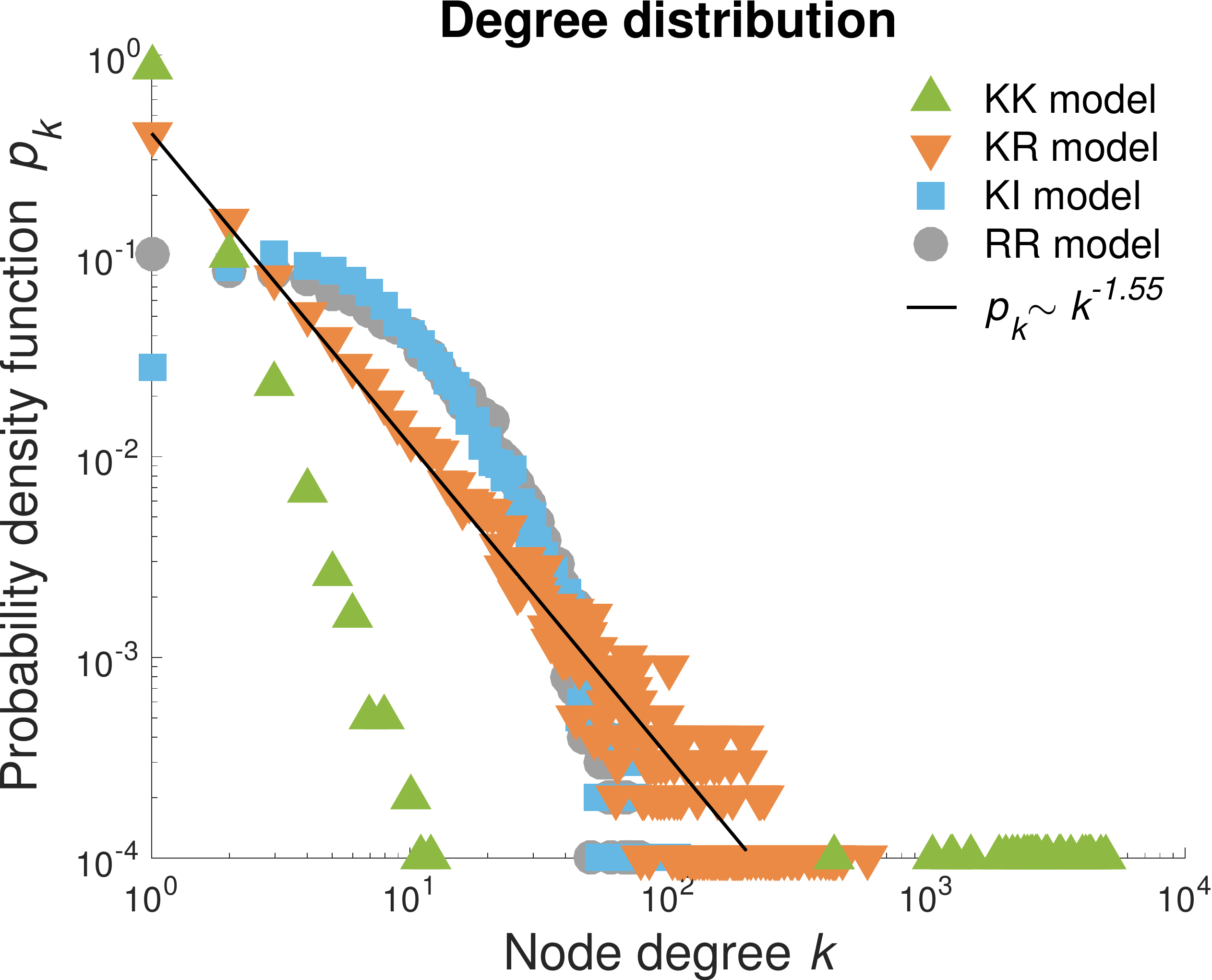}\hskip0.06\linewidth%
	\includegraphics[width=0.29\linewidth]{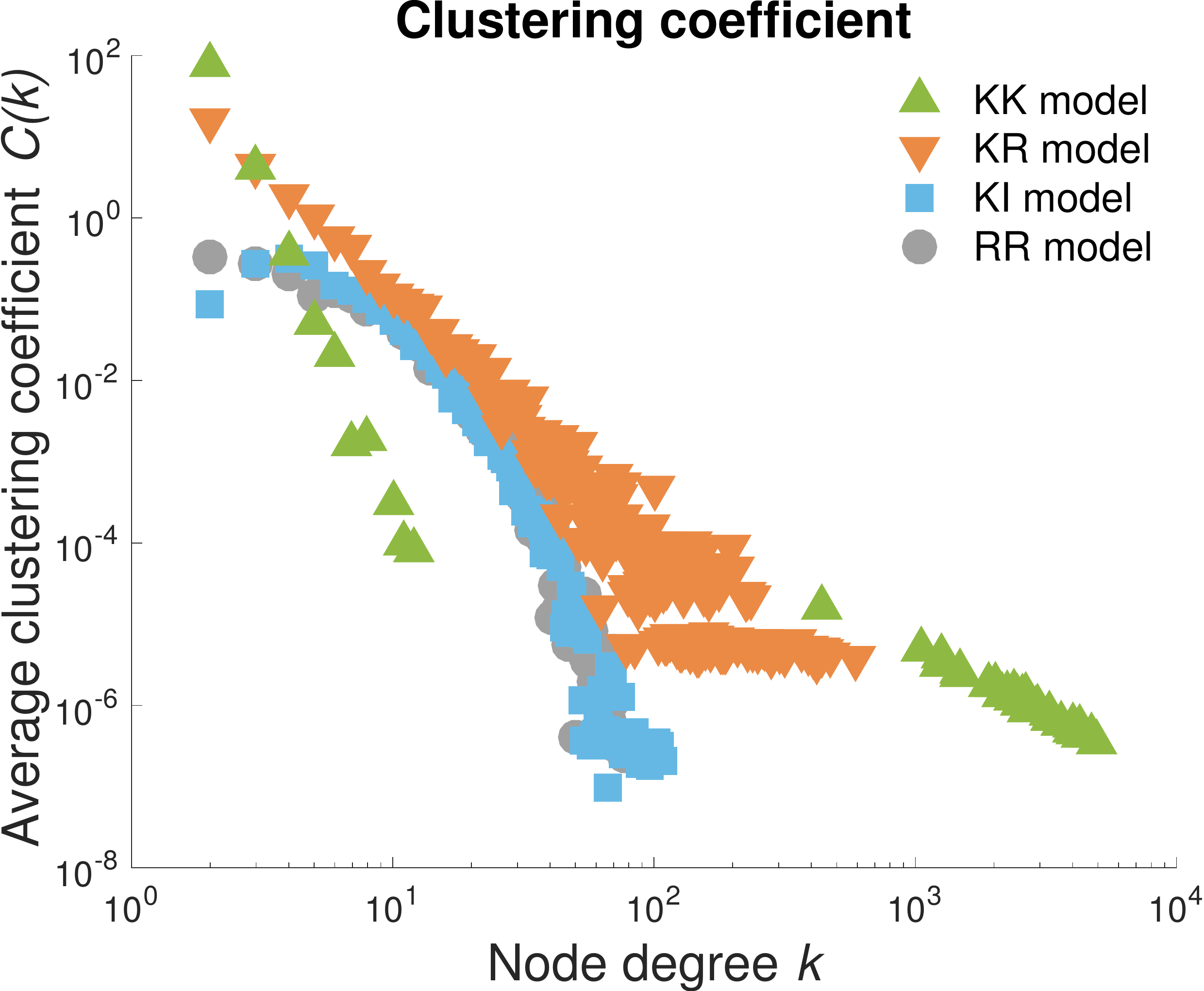}\hskip0.06\linewidth%
	\includegraphics[width=0.29\linewidth]{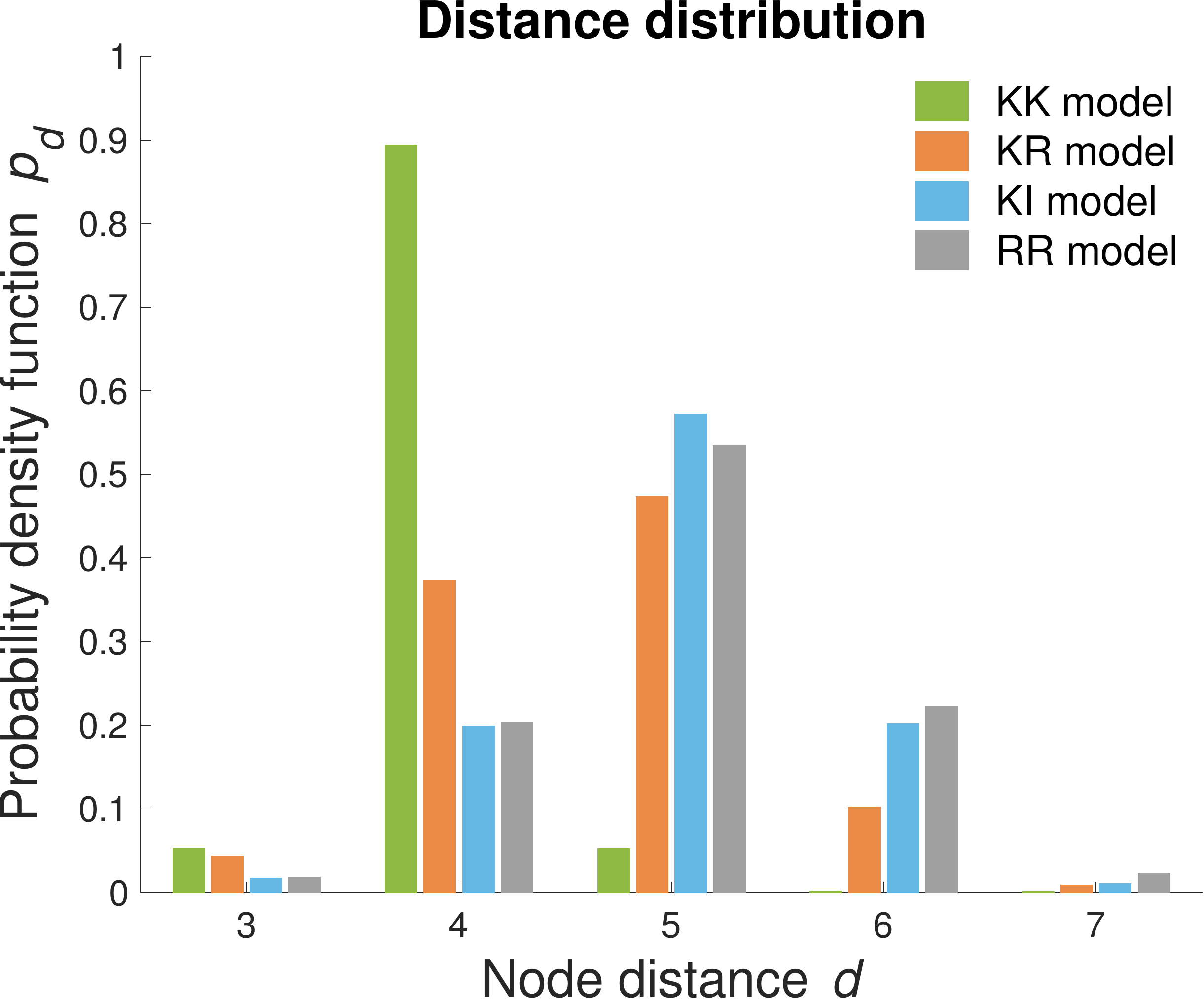}\\\vskip0.001\linewidth%
	\includegraphics[width=0.29\linewidth]{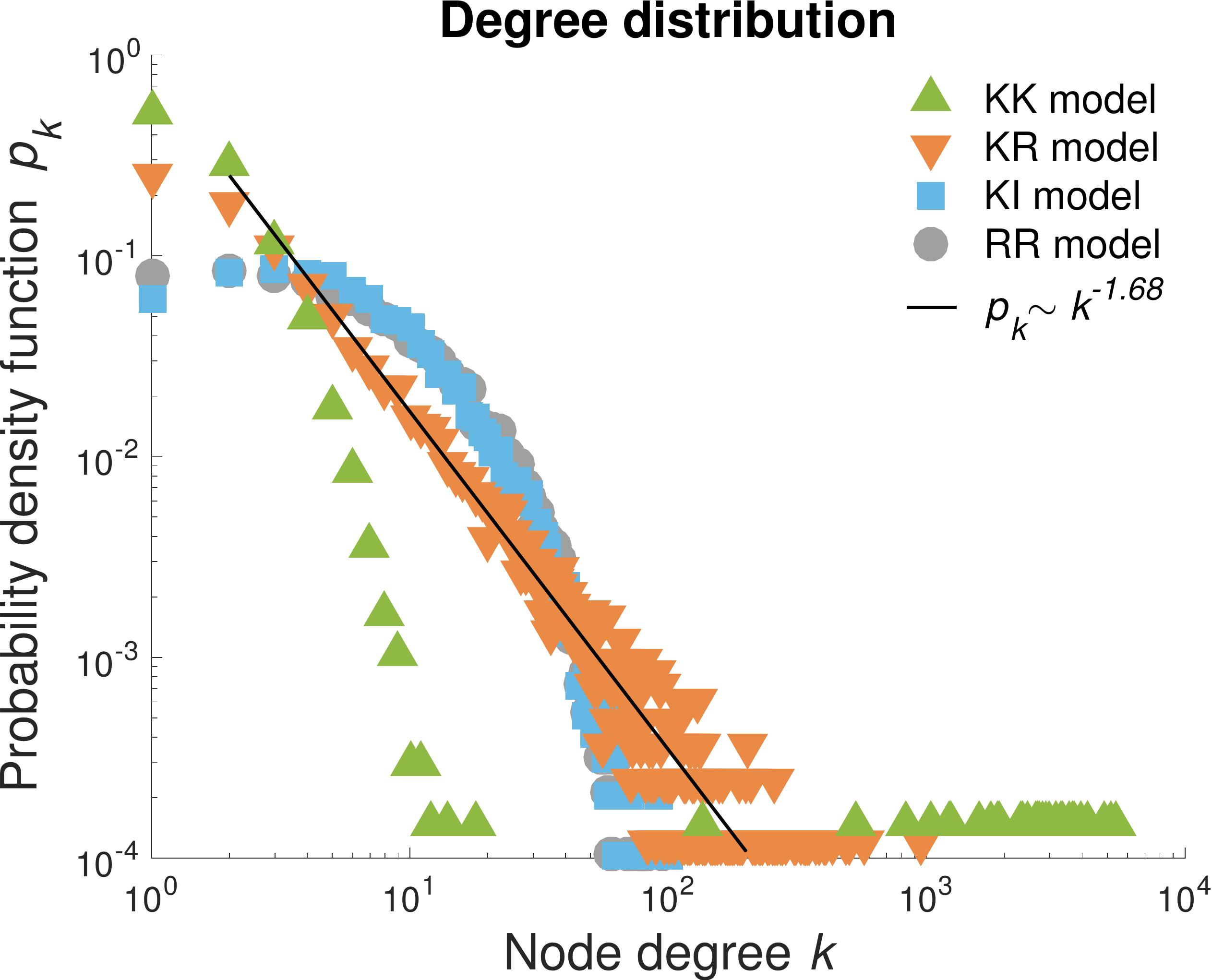}\hskip0.06\linewidth%
	\includegraphics[width=0.29\linewidth]{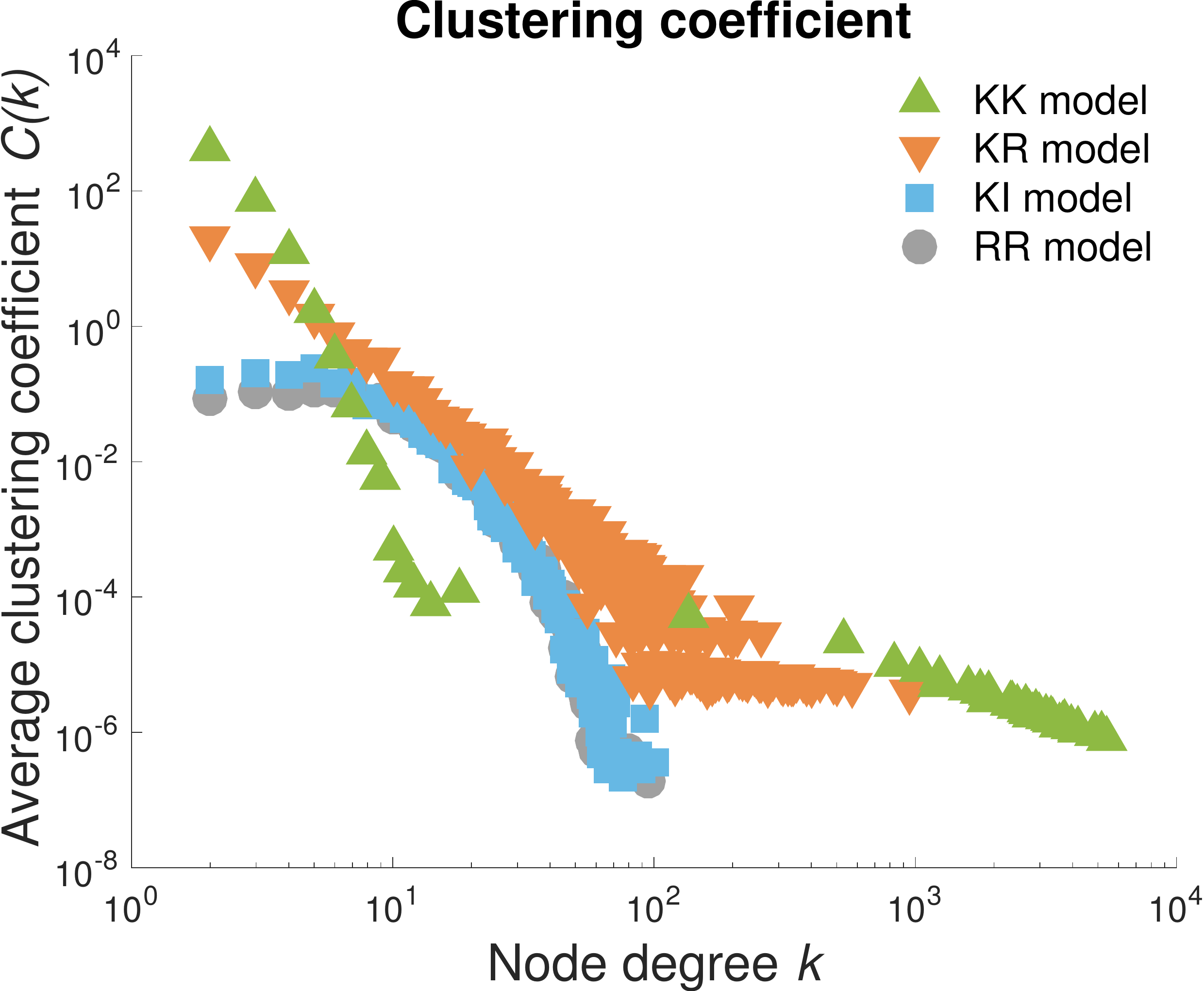}\hskip0.06\linewidth%
	\includegraphics[width=0.29\linewidth]{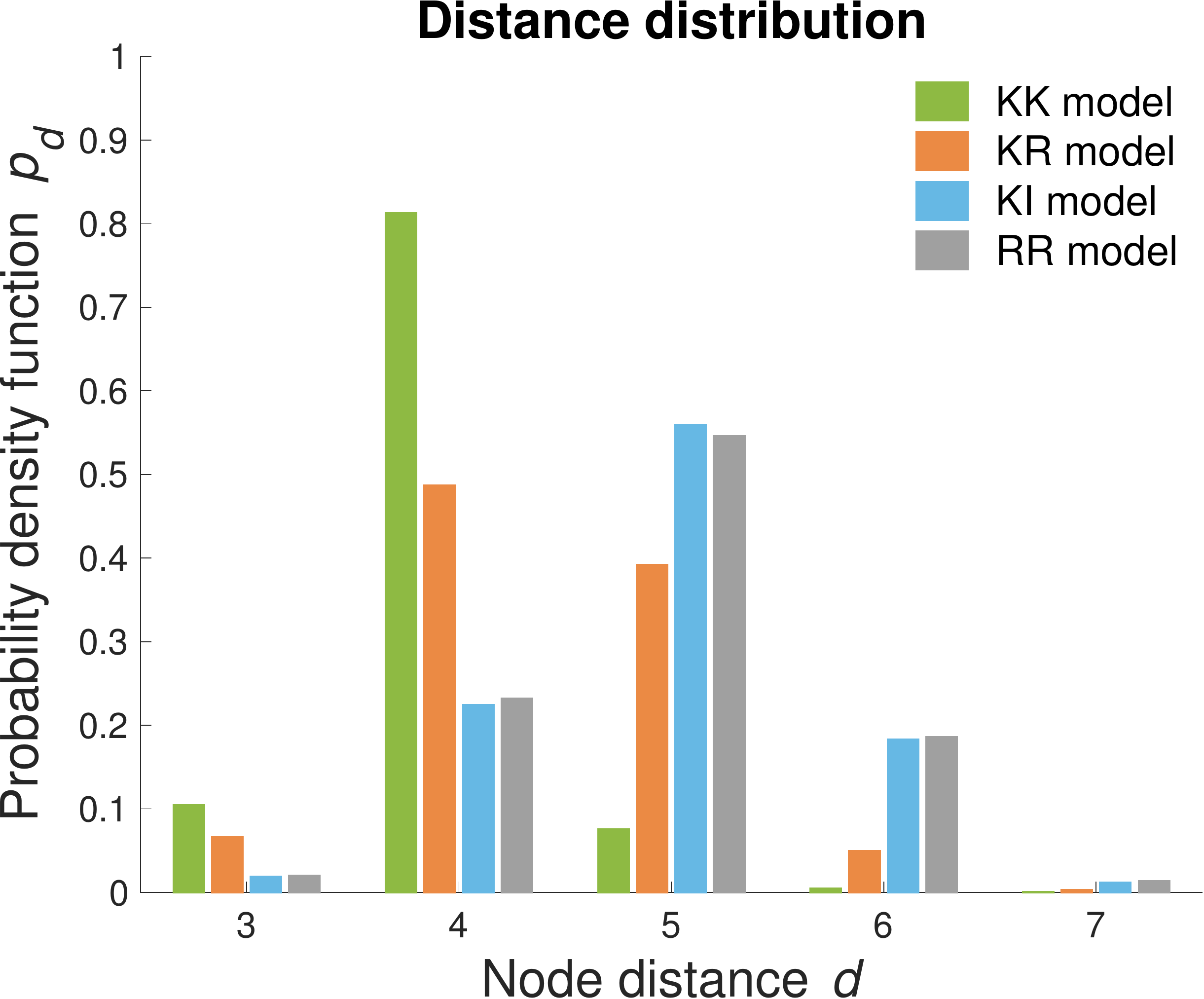}\\\vskip0.001\linewidth%
	\includegraphics[width=0.29\linewidth]{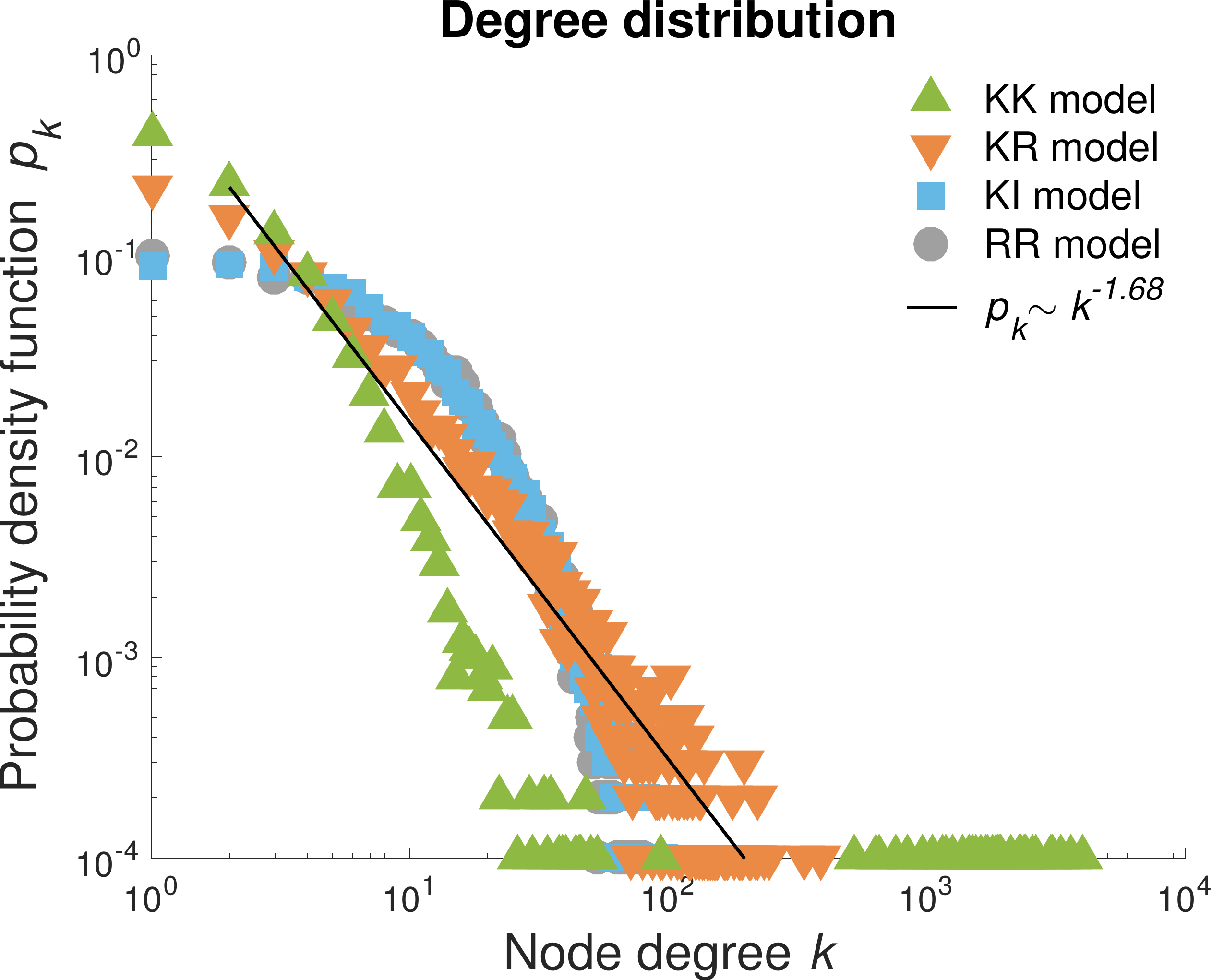}\hskip0.06\linewidth%
	\includegraphics[width=0.29\linewidth]{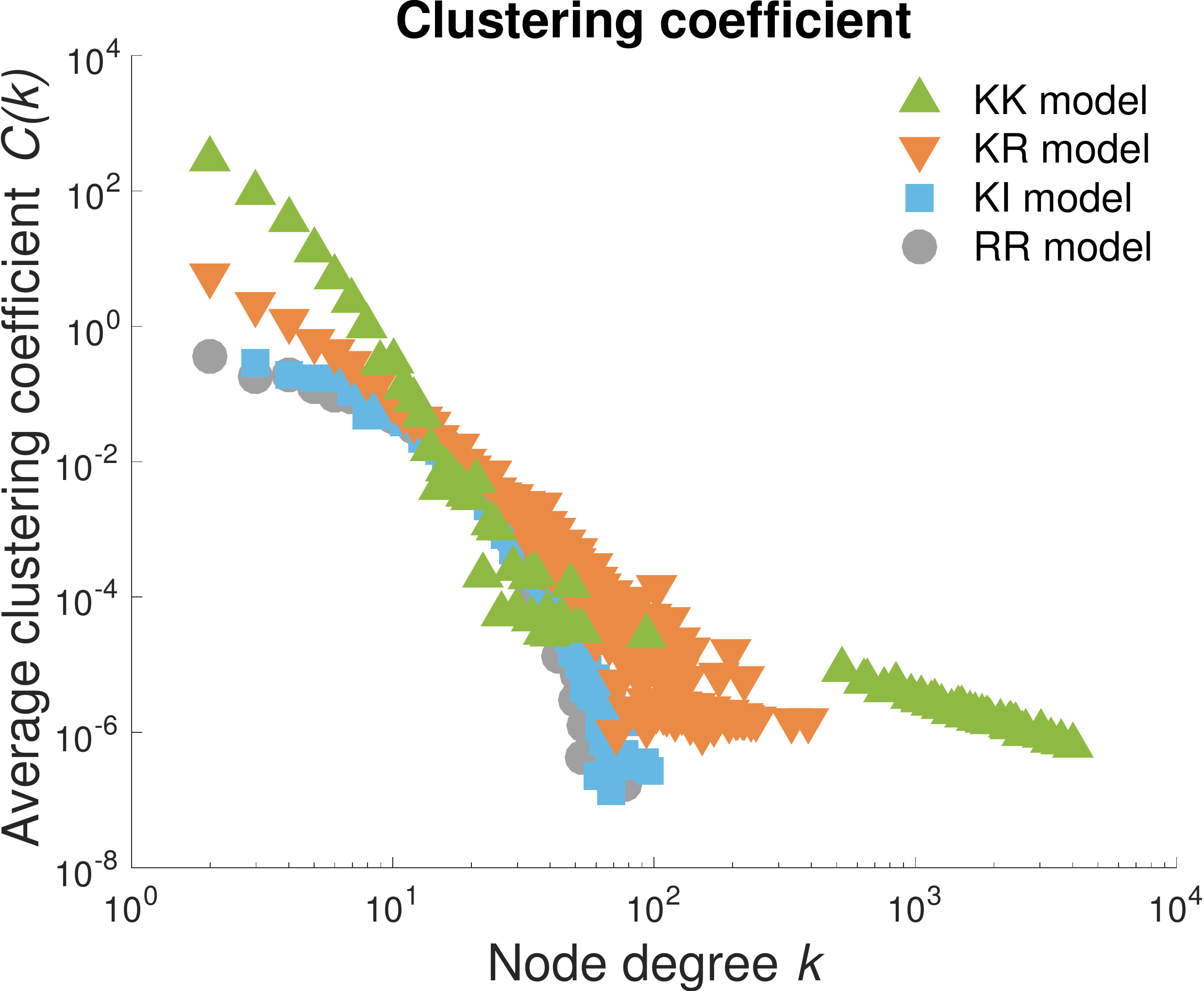}\hskip0.06\linewidth%
	\includegraphics[width=0.29\linewidth]{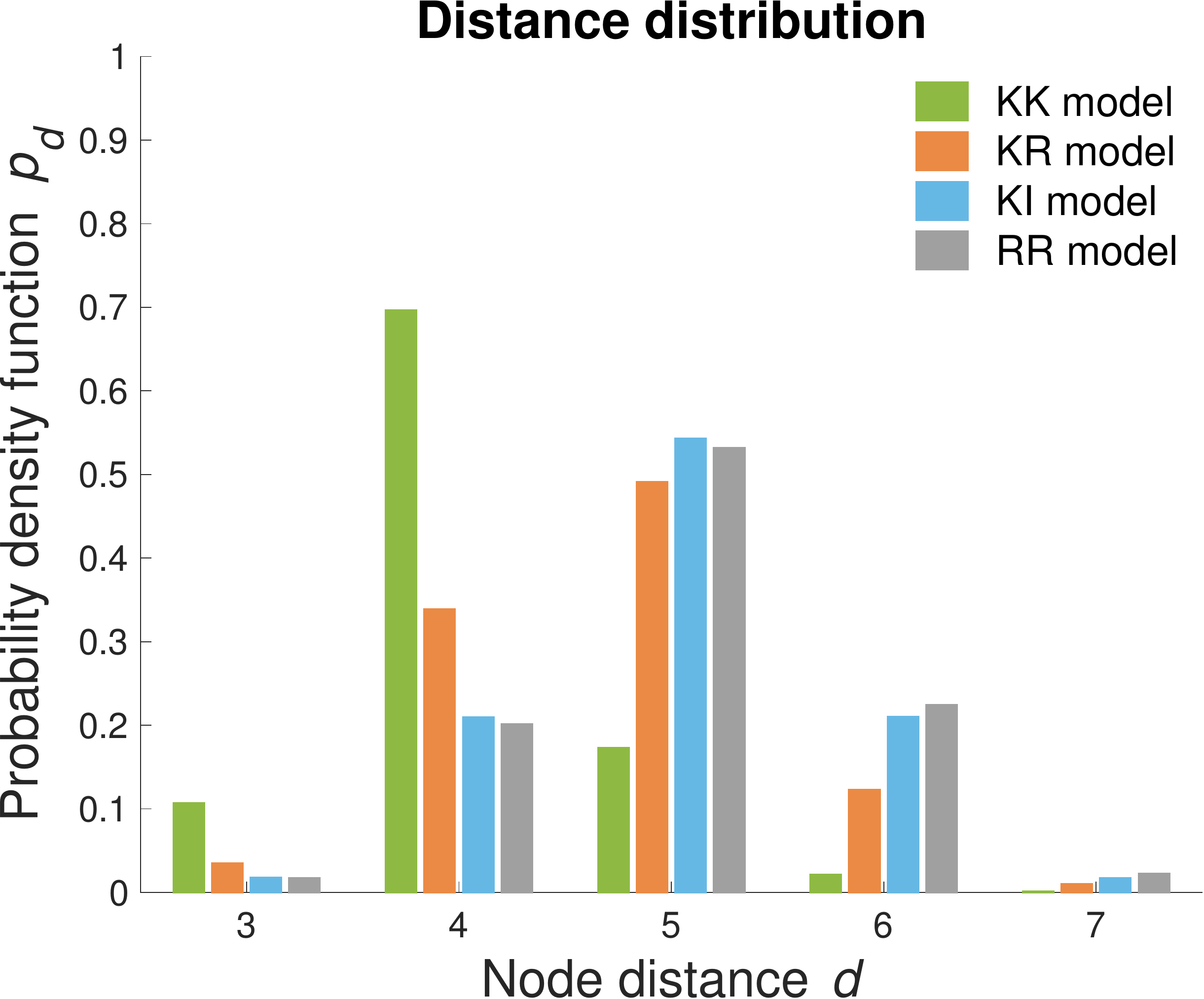}\\%
	\caption{\hlc{\bf Distributions of war pact networks.} Node degree distributions $p_k$, the average node clustering coefficient $C(k)$ and node distance distributions \hlc{$p_d$ for $d>2$} of particular realizations of the war pact model networks with $n=10\,000$ nodes and \hlc{the average degree $\mean{k}=10$}~\cite{LJTBH11}. The models are initialized either with \hlc{a perfect matching} (\emph{top}), corresponding Erd\H{o}s-R\'{e}nyi random graphs \hlc{(\emph{middle}) or a randomly grown tree graphs (\emph{bottom}).} The power-law node degree distributions $p_k\sim k^{-\gamma}$ are estimated using the maximum likelihood approach~\cite{CSN09}.}
	\label{fig:distributions}
\end{figure}

\figref{distributions} shows \hlc{distributions of various node statistics} of particular realizations of the war pact model networks. We consider four variants of the node selection rule introduced in the~\metref section and \hlc{three} different choices of model initialization. The top row in~\figref{distributions} shows the distributions for networks initialized with \hlc{a perfect matching of nodes} as in~\algref{model}, the networks in the \hlc{middle} row are initialized with Erd\H{o}s-R\'{e}nyi random graphs~\cite{ER59} with the same number of nodes and edges, \hlc{while the networks in the bottom row are initialized with randomly grown tree graphs}.

Notice that the distributions in the top, \hlc{middle} and bottom rows are almost indistinguishable. Hence, the particular choice of the model initialization has no apparent effect on the structure of the generated networks. In the \hlg{remainder}, we therefore always initialize the model with \hlc{a perfect matching of nodes}.

In contrast, the choice of the node selection rule does indeed shape the structure of the generated networks as already observed in~\figref{layouts}. For instance, consider the node degree distributions $p_k$ shown in the first column in~\figref{distributions}. When both nodes to be merged are selected preferentially according to their degree $k$ (KK model), the degree distribution $p_k$ seems to follow a power-law for low degrees $k\lesssim 10$, whereas high degree nodes $k\gtrsim 1\,000$ form a rich club~\cite{ZM04}. \hlc{Actually, the subgraph induced by the nodes with degree $k\geq 1\,000$ is a clique.} Next, when selecting the second node uniformly at random (KR model), \hlc{the degree distribution has a shape close to the power-law $p_k\sim k^{-\gamma}$} with $\gamma\approx 1.6$ throughout the entire range of node degrees. The war pact model can therefore generate scale-free networks as \hlg{is} commonly observed in social and information domains~\cite{BC19,VHHK18}. Finally, the other two node selection rules (KI and RR models) generate networks with a peak in the degree distribution characteristic of technological networks and random graphs. Hence, depending on the particular real network being modeled, different node selection rules prove appropriate.

The middle column in~\figref{distributions} shows the distributions of the average node clustering coefficient $C(k)$ for nodes with degree $k$. These largely resemble the node degree distributions $p_k$. In the case of the KR model networks with \hlc{a seemingly power-law degree distribution $p_k\sim k^{-\gamma}$}, $C(k)$ distributions also \hlc{seem to follow a power-law}~\cite{SV05}. More importantly, in all cases considered, the war pact model generates networks with \hlg{a non-trivial} node clustering coefficient $\mean{C}\gg 0$ characteristic of small-world networks~\cite{WS98}.

The small-world networks are further characterized by short distances between the nodes~\cite{WS98}. The last column in~\figref{distributions} shows the distributions of node distances \hlc{$p_d$ for $d>2$}. Most pairs of nodes are at distance $d=4$ or $5$ regardless of the particular variant of the model. Thus, in summary, the war pact model generates networks with \hlg{a scale-free} and small-world structure as commonly observed in practice.

\begin{figure}[t]
	\includegraphics[width=0.29\linewidth]{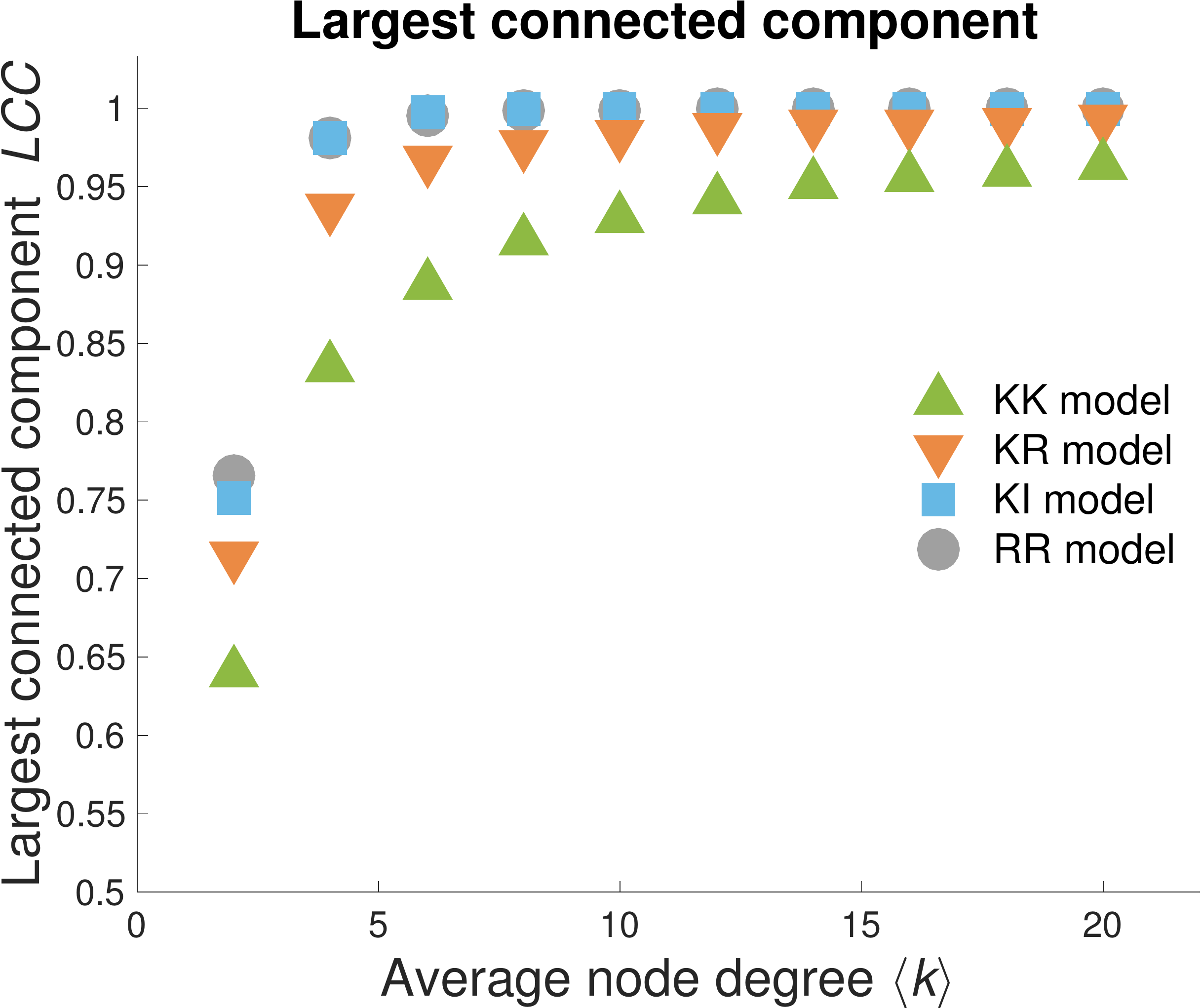}\hskip0.06\linewidth%
	\includegraphics[width=0.29\linewidth]{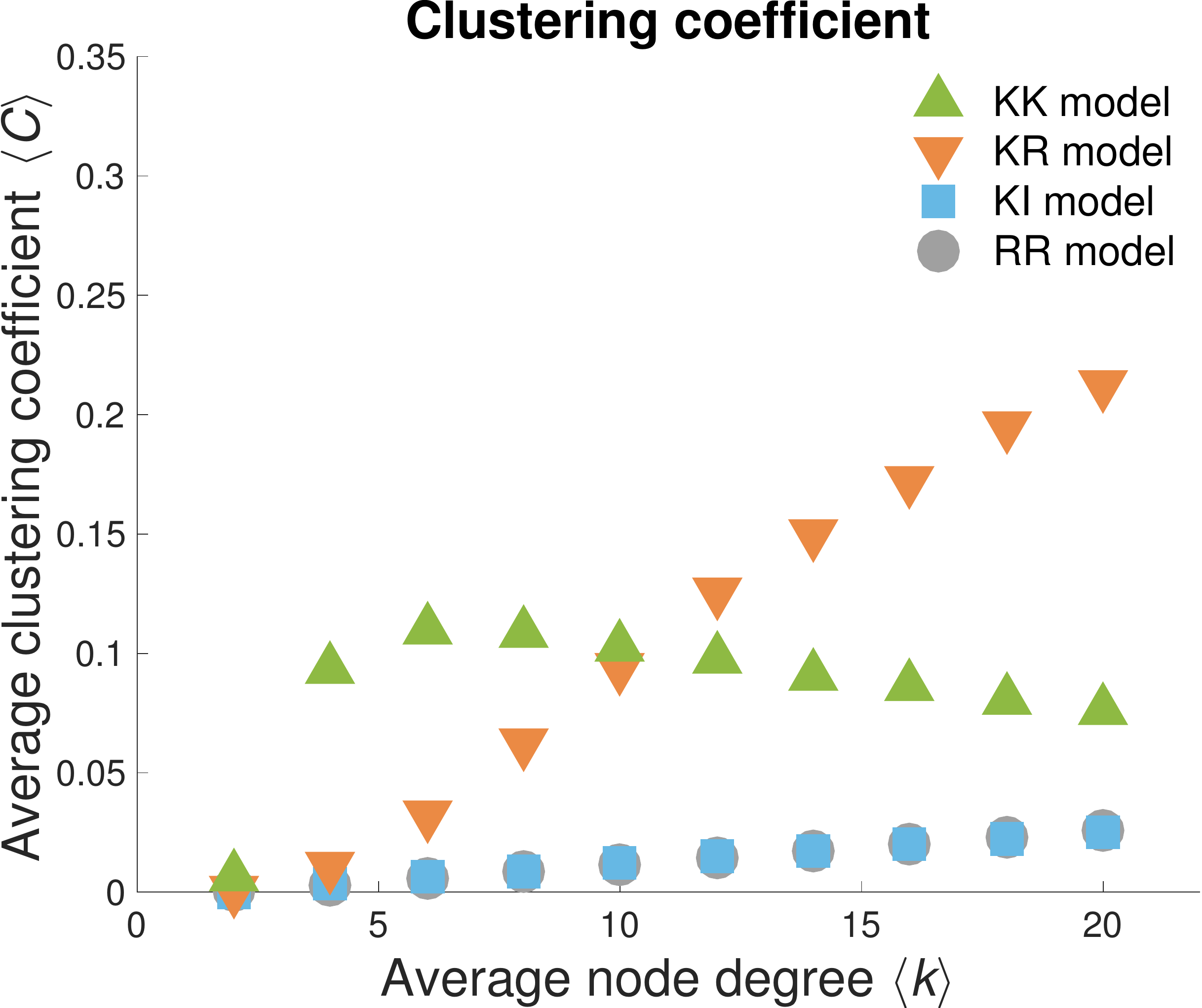}\hskip0.06\linewidth%
	\includegraphics[width=0.29\linewidth]{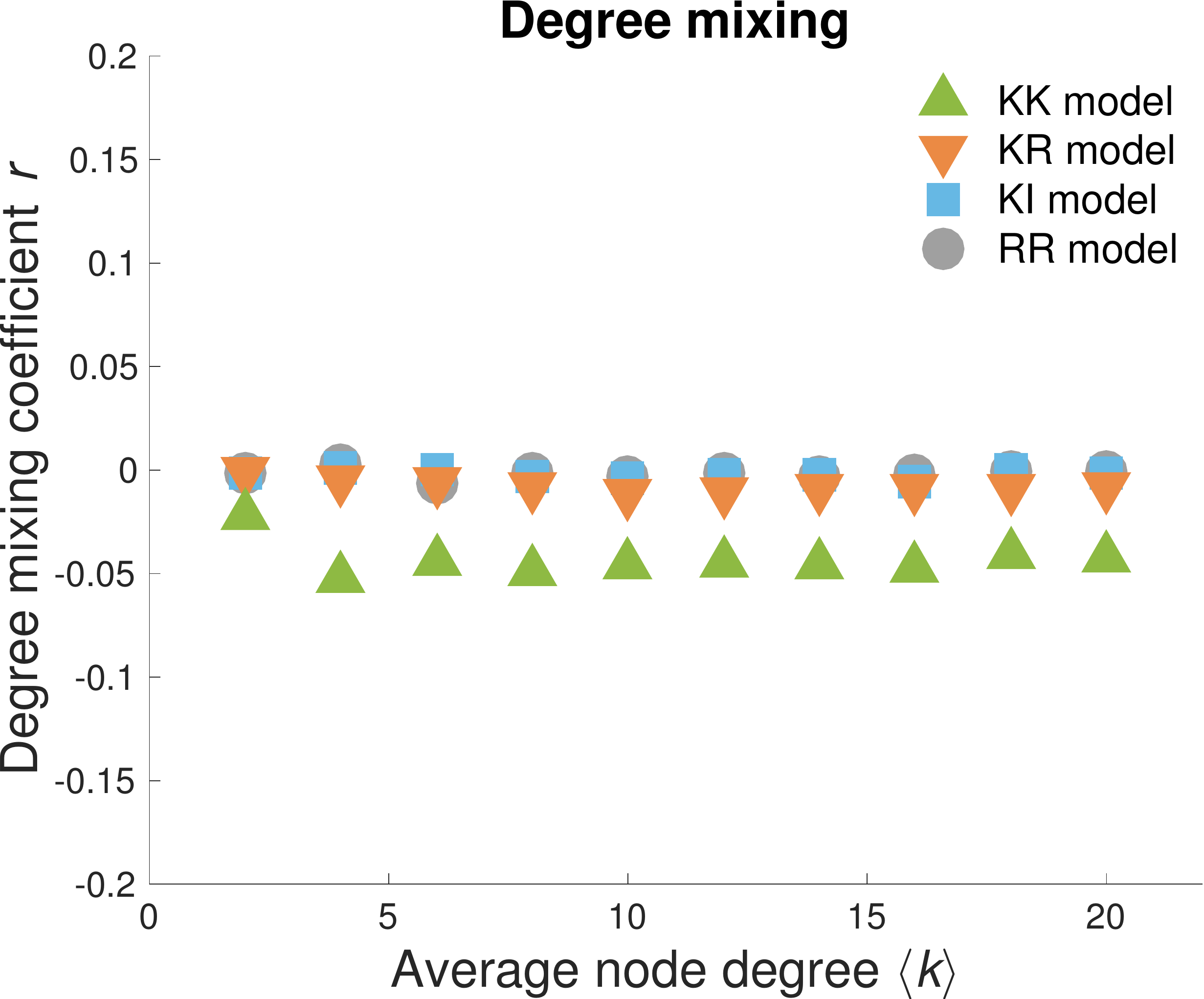}\\\vskip0.001\linewidth%
	\includegraphics[width=0.29\linewidth]{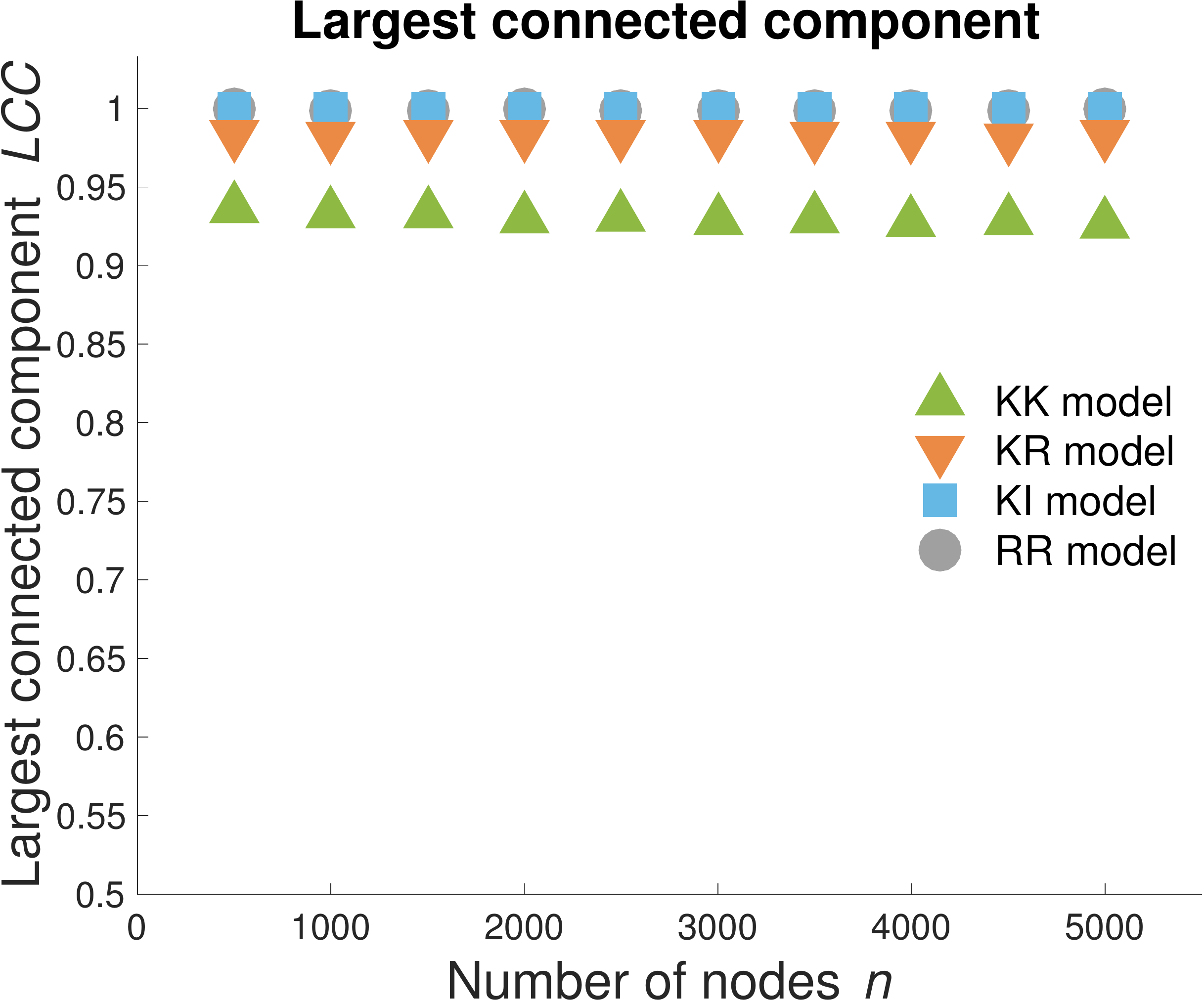}\hskip0.06\linewidth%
	\includegraphics[width=0.29\linewidth]{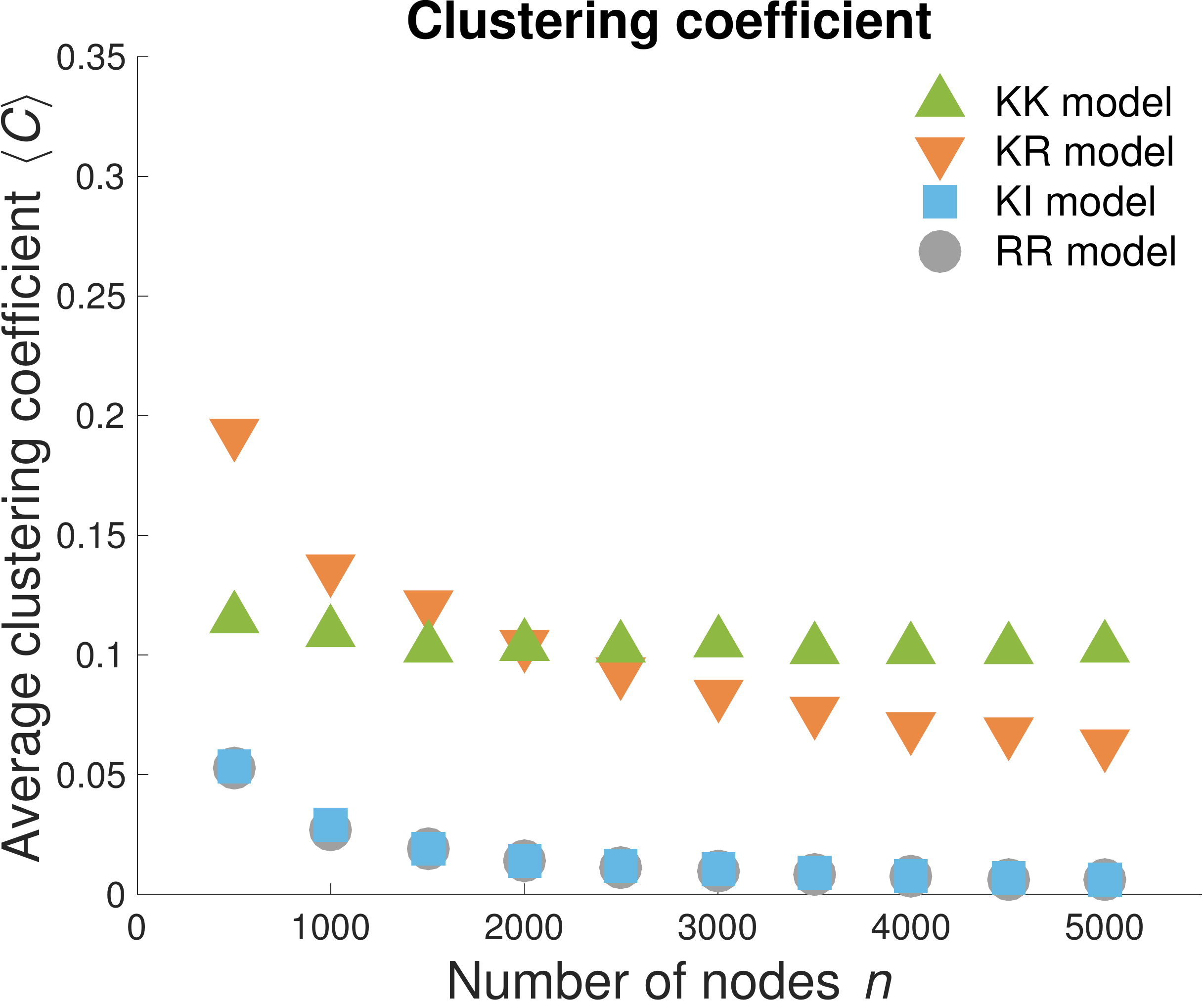}\hskip0.06\linewidth%
	\includegraphics[width=0.29\linewidth]{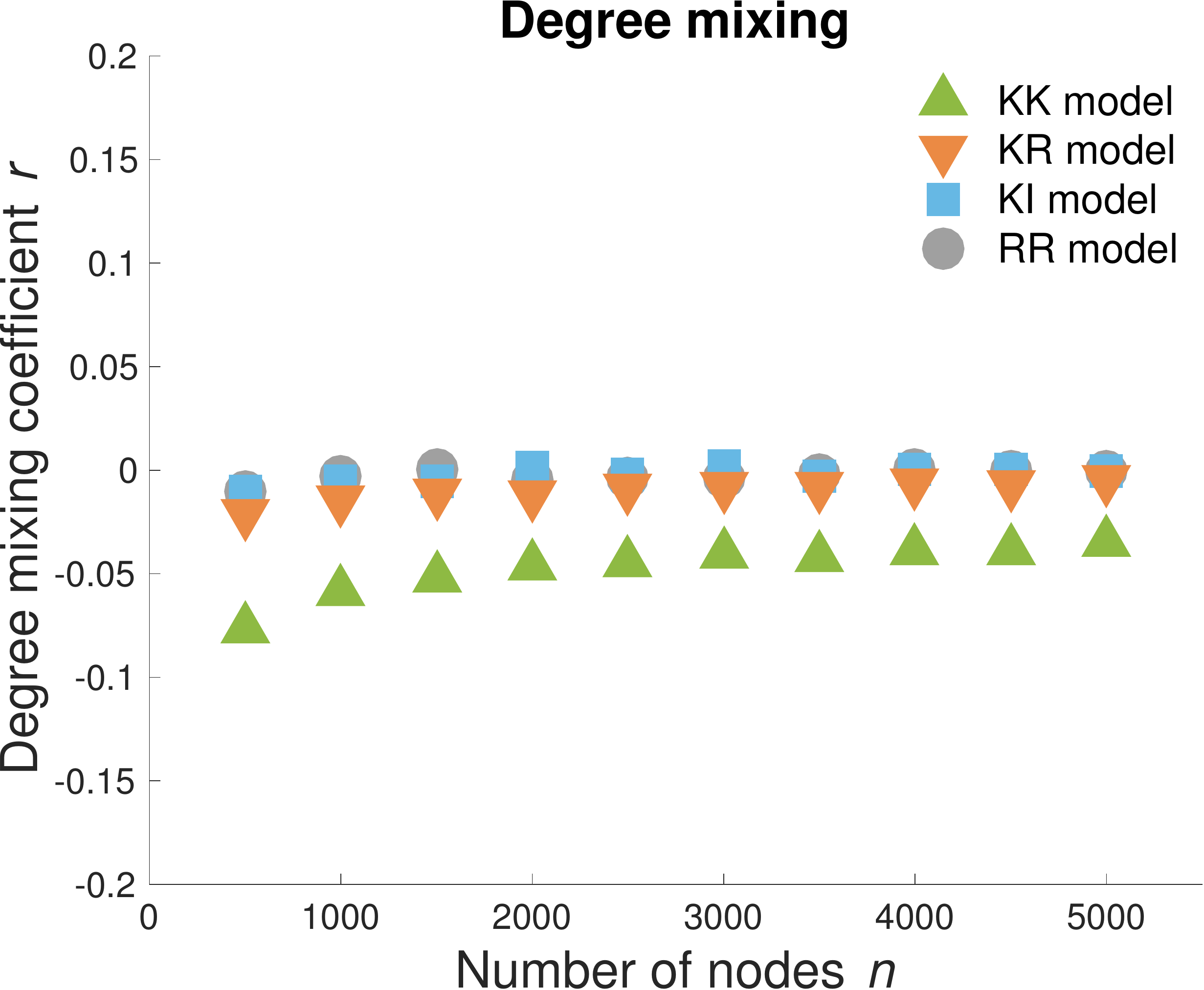}\\%
	\caption{{\bf Evolution of war pact networks.} The fractions of nodes in the largest connected component \emph{LCC}, the average node clustering coefficient $\mean{C}$ and the node degree mixing coefficients $r$ during the evolution of the war pact model networks with $n=2\,500$ nodes and growing average degree $\mean{k}$ (\emph{top}) or growing number of nodes $n$ and the average degree $\mean{k}=10$ (\emph{bottom}). \hlc{Therefore, the number of edges $m$ is increasing from left to right in all plots that} show the averages over $25$ independent realizations of the models.}
	\label{fig:evolution}
\end{figure}

\figref{evolution} shows different properties of the war pact model networks with \hlg{a growing} number of nodes or edges. These are the fractions of nodes in the largest connected component \emph{LCC}, the average node clustering coefficient $\mean{C}$ and the node degree mixing coefficients $r$. As predicted by \hlg{the} percolation theory for random graphs~\cite{New18c}, a \hlc{large connected component} \emph{LCC} $\approx 100\%$ emerges when the average node degree $\mean{k}$ exceeds a certain threshold, which depends on the particular variant of the model (top left plot in~\figref{evolution}). Nevertheless, when the average node degree equals $\mean{k}\approx 10$, the largest connected component includes \emph{LCC} $>90\%$ of the nodes regardless of the model considered. Notice that this is independent of the number of nodes $n$ (bottom left plot in~\figref{evolution}).

As expected, the average node clustering coefficient $\mean{C}$ increases with the average node degree $\mean{k}$ (top middle plot in~\figref{evolution}). Networks with the highest clustering coefficient $\mean{C}$ are generated by \hlg{the} KR model with values similar to those observed in real networks (see~\tblref{statistics}). In contrast, networks generated by \hlg{the} KK model show \hlg{an increasing} clustering coefficient $\mean{C}$ only up to a certain point when the average node degree equals $\mean{k}\approx 5$, after which $\mean{C}$ starts to decrease. The reason for this is that the networks start forming a \hlg{well-pronounced} rich club of a few \hlg{high-degree} nodes with $C=1$, whereas most of the nodes are pendant nodes with $C=0$. Finally, when fixing the average node degree to $\mean{k}=10$ and increasing the number of nodes $n$, \hlg{the} clustering coefficient $\mean{C}$ decreases for all variants of the war pact model since the generated networks are becoming increasingly more sparse (bottom middle plot in~\figref{evolution}).

The last column in~\figref{evolution} shows the evolution of the node degree mixing coefficient $r$ for the growing war pact model networks. Notice that the values of $r$ are largely independent of the number of nodes $n$ and the average node degree $\mean{k}$. All variants of the war pact model except \hlc{maybe} \hlg{the} KK model generate networks with no pronounced degree mixing $r\approx 0$. On the other hand, \hlg{the} KK model networks are \hlc{very mildly} degree disassortative with $r\approx -0.05$, due to the reasons already mentioned above.

\subsection*{Comparison and discussion}

The previous subsection shows that the choice of the war pact model initialization does not have any apparent effect on the generated networks. On the contrary, different node selection rules do indeed generate networks with \hlg{a different} topological structure. Most realistic networks matching the properties of connected scale-free and small-world networks with \hlg{a core-periphery} structure~\cite{WS98,BA99,BE00} seem to be generated by the model that selects the first node preferentially according to its current degree and the second node uniformly at random (\hlg{the} KR model). \hlc{This model is also theoretically the most sound, since it incorporates the important realism of real networks known as preferential attachment}~\cite{BA99}, \hlc{where new nodes preferentially link to well-connected nodes. Here the nodes are not added but merged, while the former are modeled by randomly selected nodes and the latter are modeled by high-degree nodes.} In the present subsection, we \hlg{also} evaluate this hypothesis empirically using four real networks from diverse domains.

\begin{figure}[t]
	\includegraphics[width=0.33\linewidth]{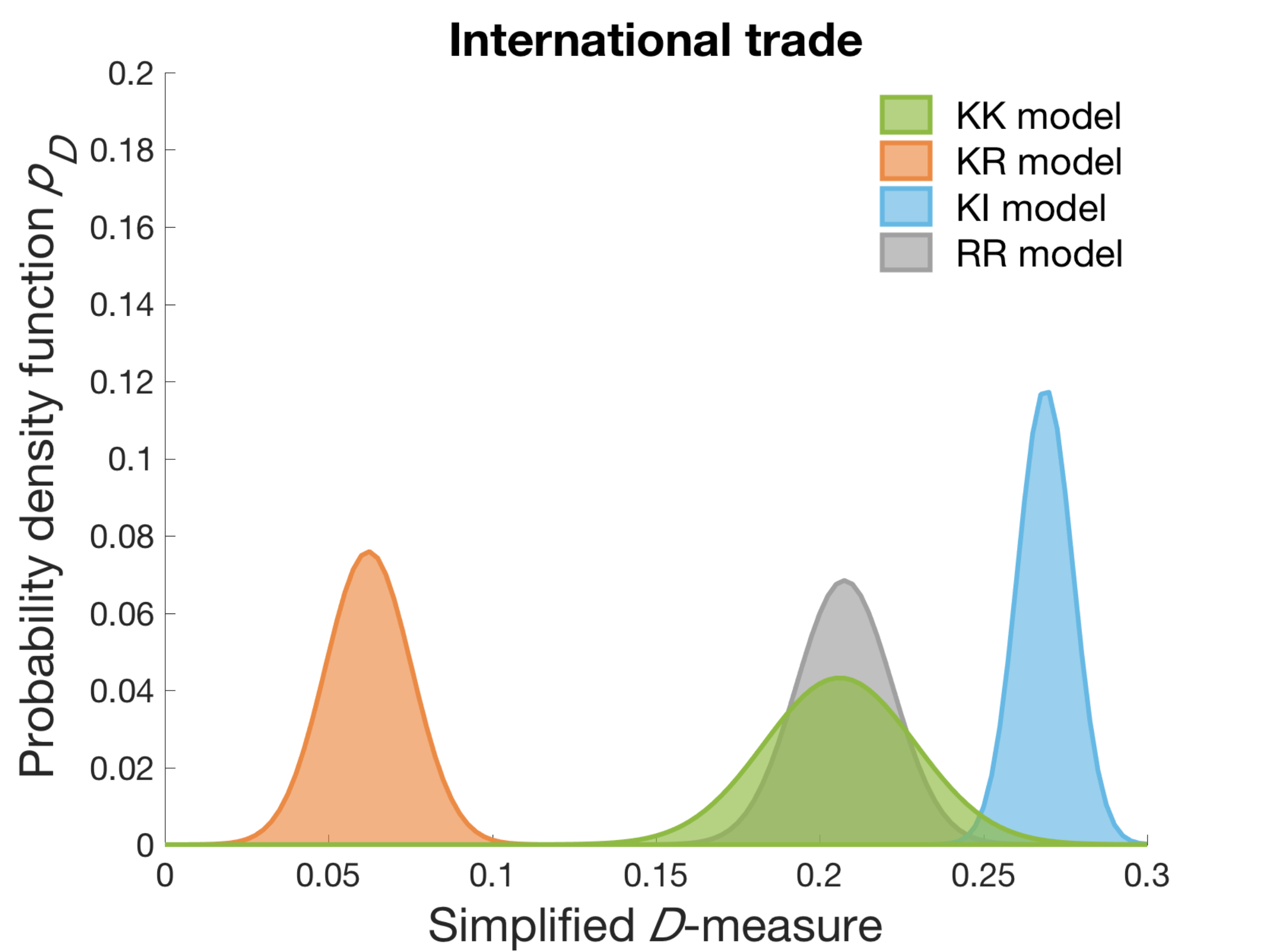}%
	\includegraphics[width=0.33\linewidth]{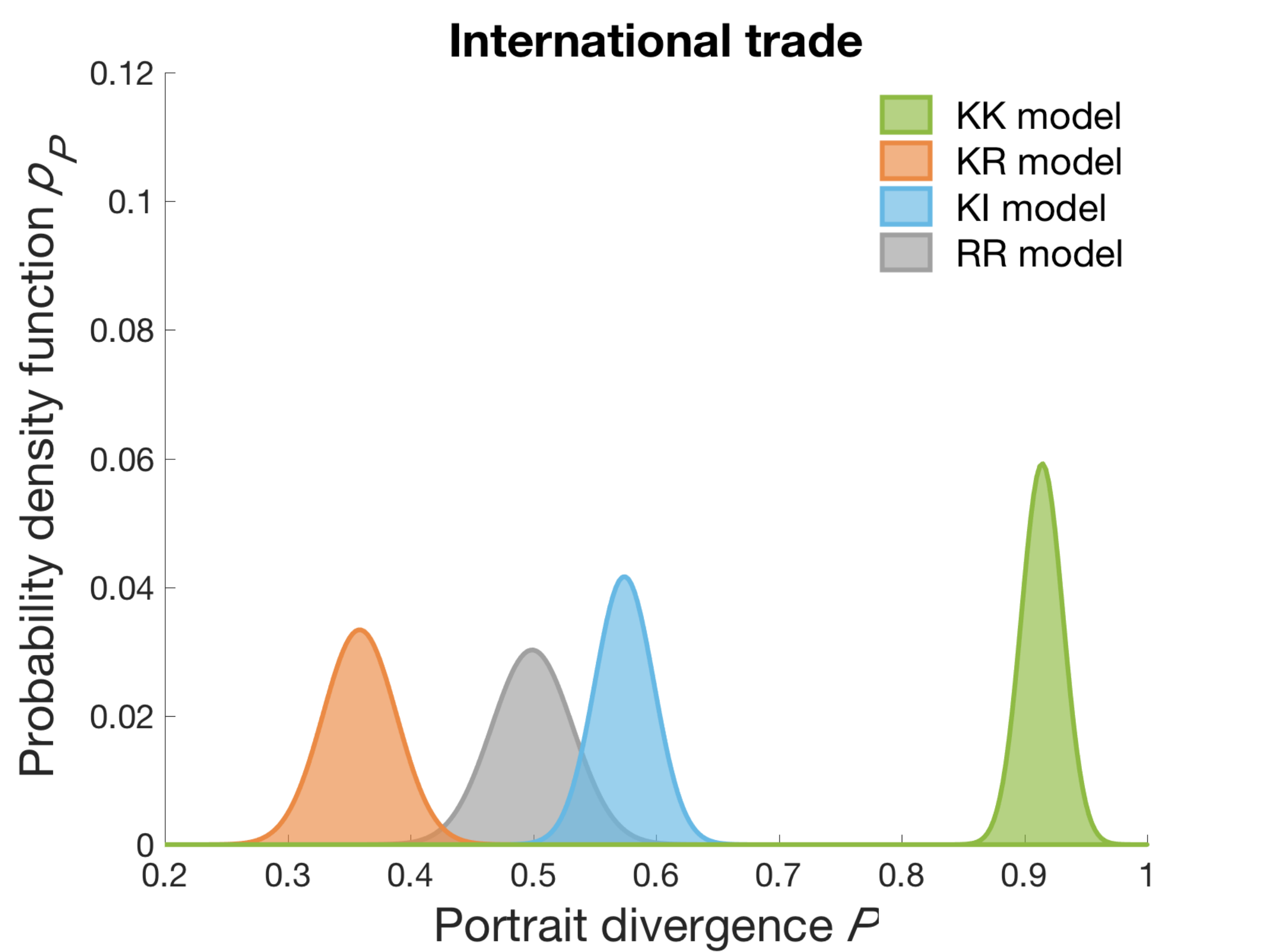}%
	\includegraphics[width=0.33\linewidth]{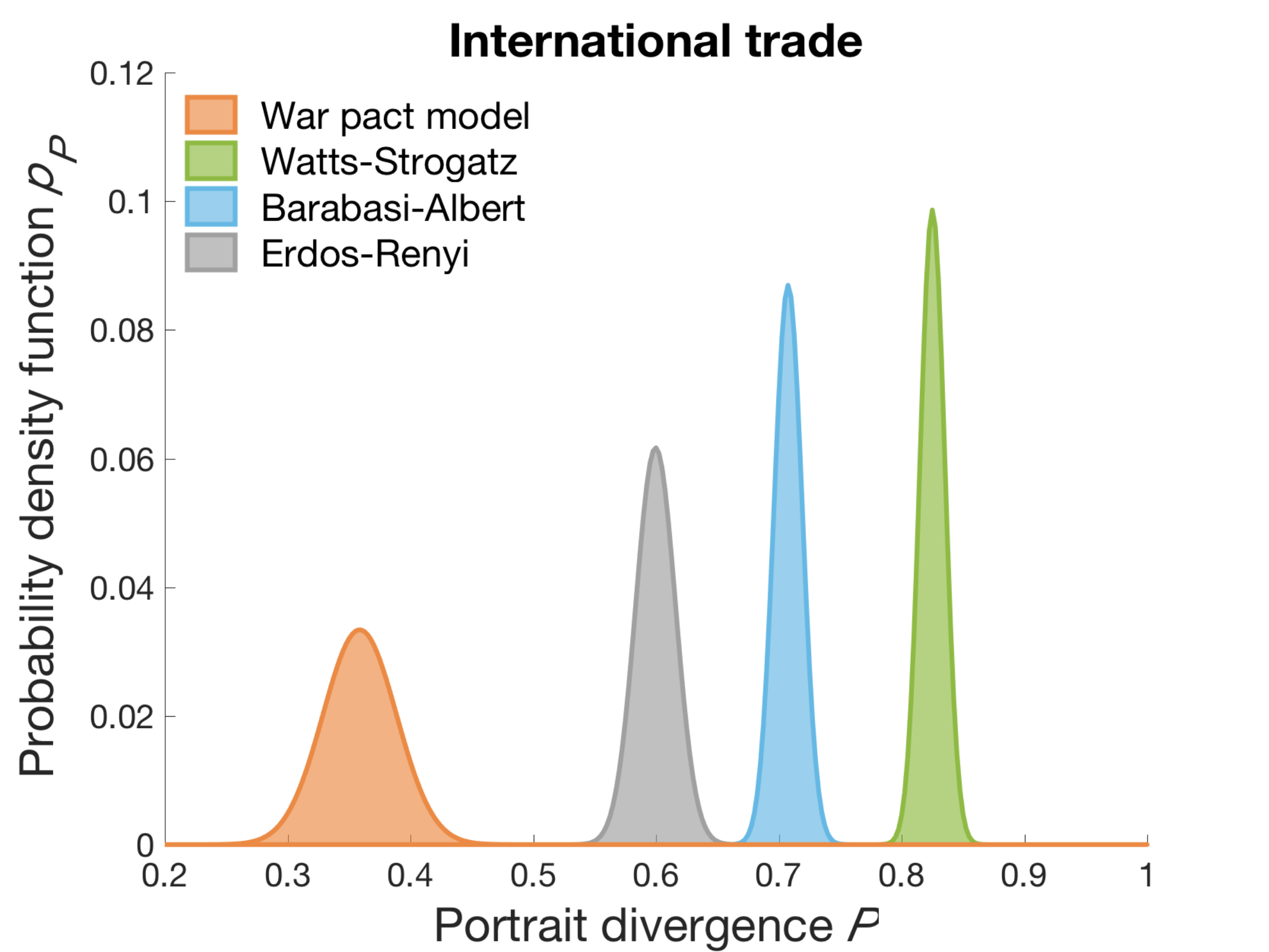}\\\vskip0.001\linewidth%
	\includegraphics[width=0.33\linewidth]{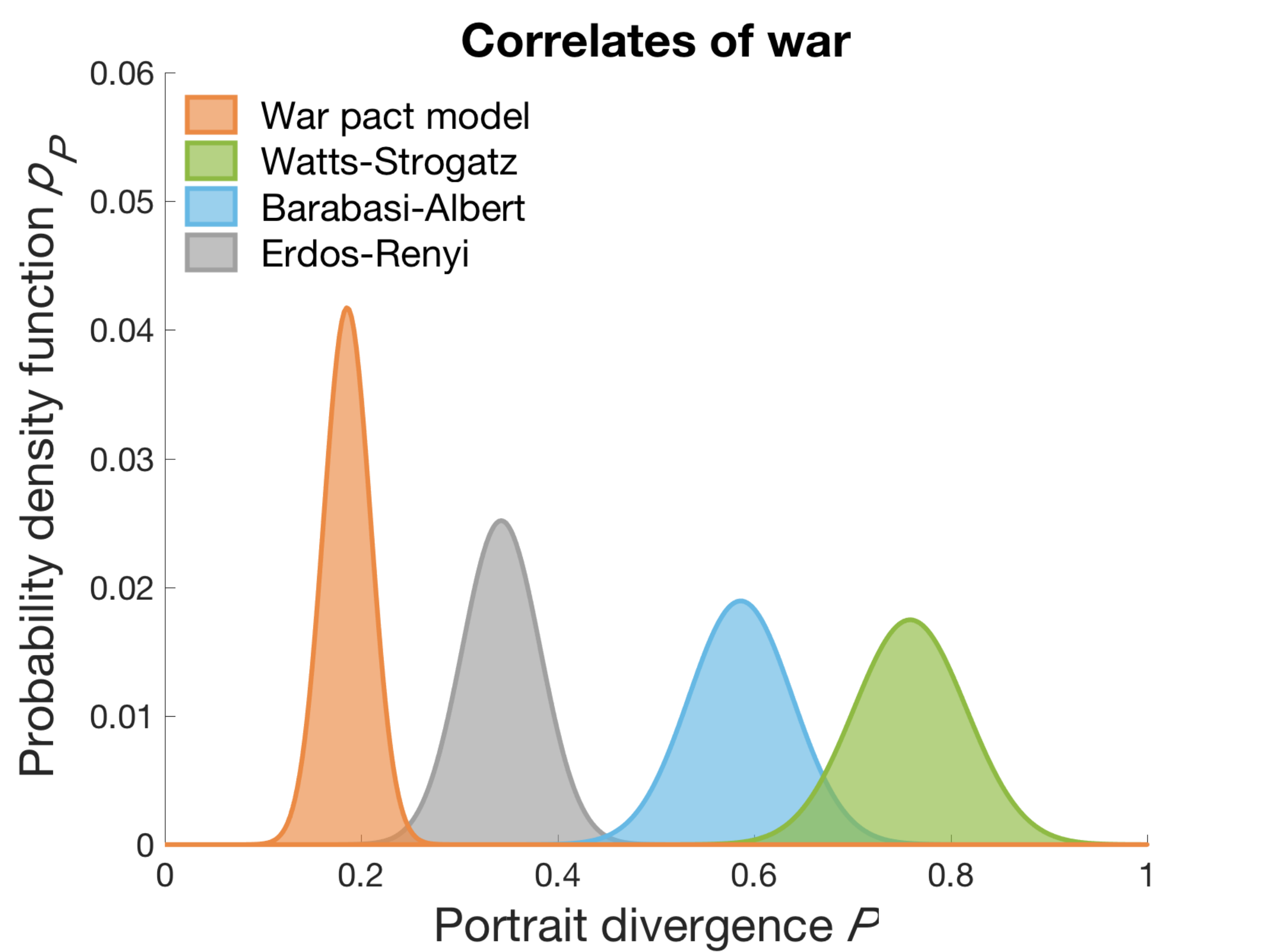}%
	\includegraphics[width=0.33\linewidth]{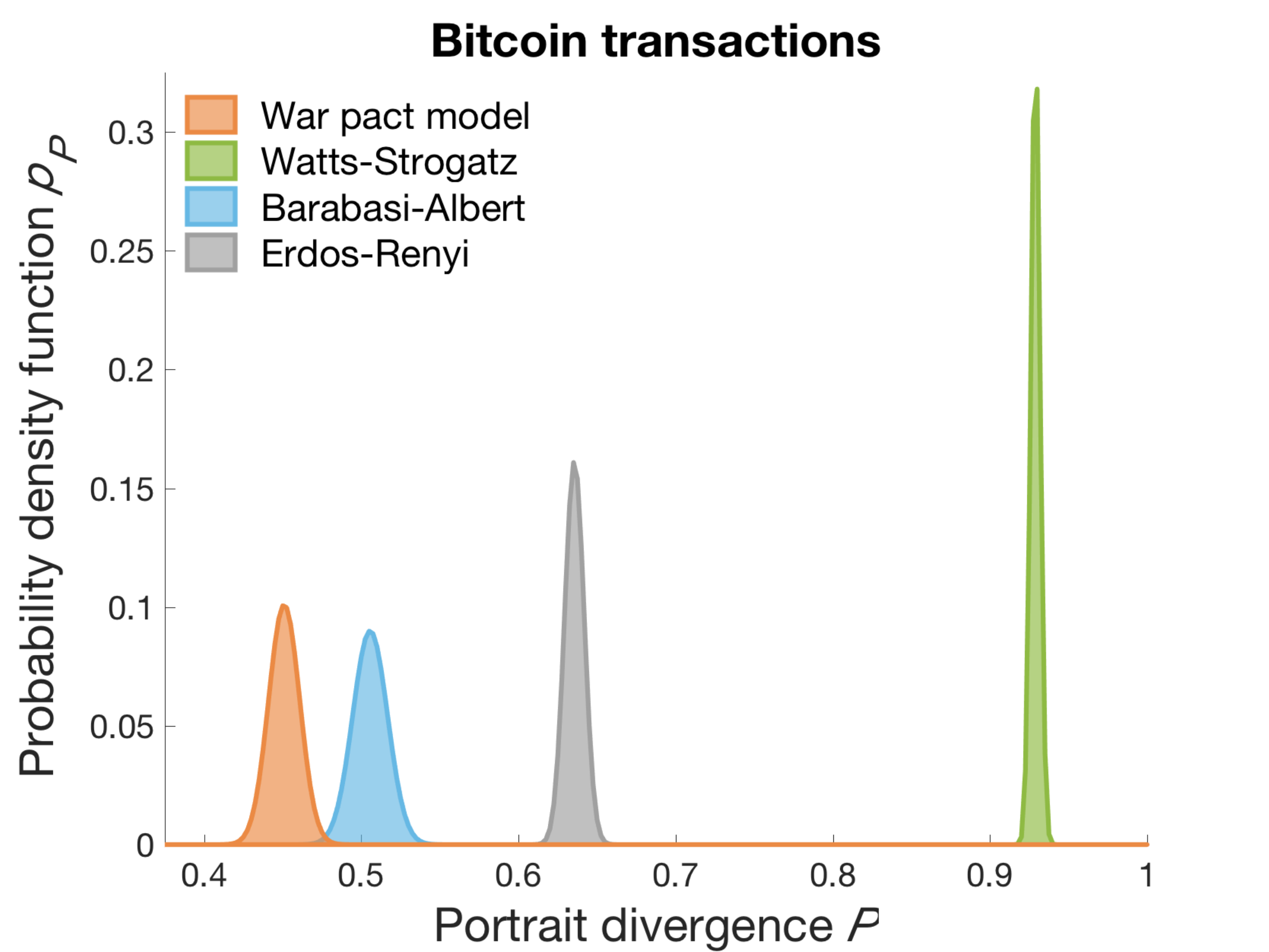}%
	\includegraphics[width=0.33\linewidth]{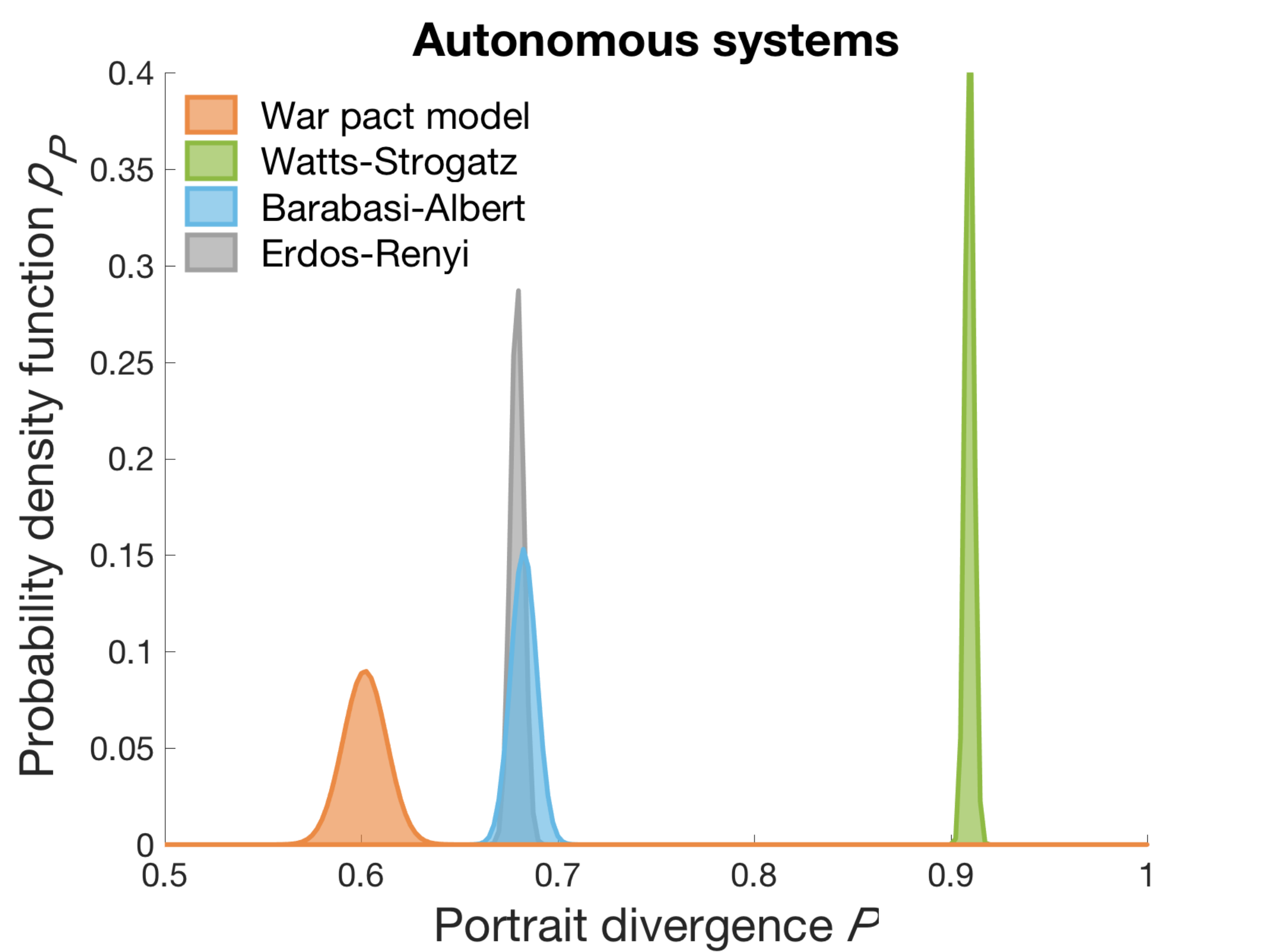}\\%
	\caption{\hlc{\bf Comparison of network models.} Comparison of the war pact model networks and classical random graphs with \hlg{the} international trade network (\emph{top}), and correlates of \hlg{the} war network, \hlg{the} Bitcoin transactions network and \hlg{the} autonomous systems graph (\emph{bottom}). The plots show distributions of the simplified $D$-measure $p_D$ and the portrait divergence $p_P$ estimated over $100$ independent realizations of the models.}
	\label{fig:comparison}
\end{figure}

The first two plots in~\figref{comparison} compare different variants of the war pact model with \hlg{an} international trade network~\cite{DNAL15}. The plots show distributions of the simplified $D$-measure $p_D$ in~\eqref{dmeasure}~\cite{SCDPMR17} and the portrait divergence $p_P$ in~\eqref{portrait}~\cite{BB18}. For both measures, \hlg{the} KR model clearly provides the best fit to the real network. Almost any network generated by \hlg{the} KR model reproduces the real network \hlg{better} than any realization of any alternative model. This \hlg{also applies to} other real networks analysed in the paper (exact results are omitted). In the \hlg{remainder}, we therefore compare other random graph models only with \hlg{the} KR model.

The remaining four plots in~\figref{comparison} compare networks generated by different random graph models with \hlg{the} international trade network as above, and correlates of \hlg{the} war network~\cite{DM15b}, \hlg{the} Bitcoin transactions network~\cite{KCSPV14} and \hlg{the} autonomous systems graph~\cite{LKF07}. The plots show the distributions of the portrait divergence $p_P$, while the models include the war pact networks, Erd\H{o}s-R\'{e}nyi random graphs~\cite{ER59}, Barab\'{a}si-Albert scale-free networks~\cite{BA99} and Watts-Strogatz small-world networks~\cite{WS98}. The latter are without doubt the most fundamental and commonly analysed models in network science literature.

\hlg{All} real networks considered, the war pact model reproduces the structure of the networks \hlg{better} than any other model. Again, almost any network generated by the war pact model fits the real network \hlg{better} than any realization of any alternative model. We stress that all these models are either static or models with \hlg{a growing} number of nodes and edges. In contrast, the war pact model networks shrink over time and possibly provide a better explanation of the evolution of the considered real networks.

\tblref{comparison} further shows the standard statistics of the war pact model networks that best reproduce real networks according to the portrait divergence. Comparing the values with those in~\tblref{statistics}, these match the statistics of real networks \hlg{well} with a few exceptions shown in bold in~\tblref{comparison}. In particular, the war pact model underestimates the average node clustering coefficient $\mean{C}$ and overestimates the node degree mixing coefficient $r$  in correlates of war and Bitcoin transaction networks, and \hlg{the} autonomous systems graph, while the model underestimates the modularity $Q$ of \hlg{the} community structure in international trade and Bitcoin transactions networks, and \hlg{the} autonomous systems graph. On the other hand, the model almost precisely reproduces the fraction of nodes in the largest connected component \emph{LCC}, the average distance between the nodes $\mean{d}$ and network diameter $d_{max}$. Overall, the war pact model replicates the structure of these real networks \hlg{better} than any other model considered.

\begin{table}[t]
	\begin{adjustwidth}{-2.25in}{0in}
	\centering
	\caption{\hlc{\bf Statistics of war pact networks.} Standard statistics of the war pact model networks that best reproduce real networks according to the portrait divergence estimated over $100$ independent realizations of the model.}\vskip8pt
	\begin{tabular}{|l+r|r|r|r|r|r|c|r|r|} \hline
		{\bf Network} & \multicolumn{1}{c|}{$n$} & \multicolumn{1}{c|}{$m$} & \multicolumn{1}{c|}{\emph{LCC}} & \multicolumn{1}{c|}{$\mean{k}$} & \multicolumn{1}{c|}{$\mean{C}$} & \multicolumn{1}{c|}{$\mean{d}$} & \multicolumn{1}{c|}{$d_{max}$} & \multicolumn{1}{c|}{$r$} & \multicolumn{1}{c|}{$Q$} \\\thickhline
		Correlates of war & \hlc{$41$} & \hlc{$54$} & \hlc{$90.2\%$} & \hlc{$2.63$} & \hlc{$\mathbf{0.06}$} & \hlc{$2.64$} & \hlc{$7$} & \hlc{$\mathbf{-0.14}$} & \hlc{$0.53$} \\\hline
		International trade & $130$ & $3\,730$ & $100.0\%$ & $57.38$ & $0.53$ & $2.17$ & $5$ & $-0.04$ & $\mathbf{0.02}$ \\\hline
		Bitcoin transactions & $1\,288$ & $6\,236$ & $98.0\%$ & $9.68$ & $\mathbf{0.13}$ & $3.08$ & $7$ & $\mathbf{-0.05}$ & $\mathbf{0.24}$ \\\hline
		Autonomous systems & $3\,213$ & $11\,248$ & $98.3\%$ & $7.00$ & $\mathbf{0.03}$ & $3.62$ & $9$ & $\mathbf{0.00}$ & $\mathbf{0.33}$ \\\hline
	\end{tabular}
	\label{tbl:comparison}
	\end{adjustwidth}
\end{table}

Besides, the war pact model also provides an intuitive explanation of the evolution of many real networks. For instance, the nodes in \hlg{the} correlates of war network represent \hlc{alliances between} world nations and the edges represent different military or non-military conflicts between them. When \hlc{nations of two alliances} form a pact, or \hlc{nations in one alliance occupy the nations of another}, the enemies of both become common enemies, which can be modeled by simply merging the corresponding nodes. Furthermore, the node selection rule that proved most suitable above suggests that larger \hlc{alliances} with larger number of enemies \hlc{form a pact} with or conquer other \hlc{alliances}. This intuition has motivated the name war pact model.

The evolution of other real networks analysed in the paper can be explained in a similar manner. The trading relations between countries or companies are shared after an alliance between two countries or a merger of two companies. Next, when a single user controls multiple Bitcoin addresses, these are likely to coappear in future transactions. Finally, when two entities that have governed their Internet traffic independently unite for whatever reason, their traffic is merged from an external point of view. Indeed, one can come up with a similar intuitive explanation of the evolution of other real networks not considered here.

As already mentioned before, the initialization of the war pact model with pairs of connected nodes is somewhat artificial in the scenarios considered. However, as we show in the empirical evaluation of the model, the particular choice of model initialization has no apparent effect on the resulting structure of the generated networks.

% % % % % % % % % % % % %
%
%			CONCLUSION
%
% % % % % % % % % % % % %

\section*{Conclusion}

In this \hlg{paper,} we propose a simple model of shrinking networks called the war pact model. The model starts with some fixed number of edges \hlc{forming a perfect matching}, and then iteratively merges the nodes until the desired number of nodes is obtained. In contrast to most network models in literature that are either \hlg{static,} representing a snapshot of a \hlg{network}~\cite{ER59,WS98,NSW01}\hlg{,} or generate networks with \hlg{a growing} number of nodes and edges~\cite{Pri76,BA99,KKRRT99}, the war pact model networks shrink and thus represent a shift in the perspective of the evolution of real networks that has been largely neglected in the past~\cite{DM02,KNB08}.

We show that networks generated by the war pact model match \hlg{the} common properties of real networks. These include the emergence of a \hlc{large connected component}~\cite{ER59}, \hlg{a scale-free} node degree distribution~\cite{BA99}, \hlg{a small-world} network structure~\cite{WS98}, \hlg{a disassortative} node degree mixing~\cite{New02}, \hlg{a distinctive} network mesoscopic structure~\cite{BE00} and selected other properties. Even more importantly, the model provides an intuitive explanation of the evolution of diverse real networks representing the worldwide trade, international wars or non-military conflicts and other disputes, cryptocurrency transactions, Internet traffic and likely many other networks not considered here. In summary, compared to classical growing network models, network shrinking possibly provides a more reasonable explanation of the evolution of at least some real networks and greater emphasis should be put on such models in the future.

There are various directions for further research. Firstly, due to the algorithmic simplicity of the war pact model, different network properties might be derived \hlg{analytically, thus} rendering numerical simulations unnecessary. Secondly, the model could be extended to other types of networks like weighted or valued and also signed networks. Similarly, the node selection rule could be easily adjusted for multimode and multiplex networks. Finally, a thorough comparison of different network models could be conducted, possibly giving a more conclusive answer whether growing or shrinking models, \hlc{or some reasonable combination of them,} explain the evolution of real networks \hlg{better}.

% % % % % % % % % % % % %
%
%			APPENDIX
%
% % % % % % % % % % % % %

\section*{Acknowledgments}
\hlc{The authors wish to thank Vladimir Batagelj for his helpful comments and suggestions on an earlier version of the paper, and Teja Goli for proofreading the final paper.}

\section*{\hlc{Funding Statement}}
This \hlc{study} has been supported by the Slovenian Research Agency under the program P5-0168 and by the European Union COST Action number CA15109. \hlc{There was no additional external funding received for this study.}

% % % % % % % % % % % % %
%
%			BIBLIOGRAPHY
%
% % % % % % % % % % % % %

% \nolinenumbers

% \bibliographystyle{plos2015}
% \bibliography{bibliography}

\end{document}